\documentclass[secnumarabic,amssymb, nobibnotes, aps, prc]{revtex4} 

\usepackage{graphicx}
\usepackage{amssymb}
\usepackage{amsmath}
\usepackage{bm}
\usepackage{kotex}
\usepackage{times}
\usepackage{xcolor}
\usepackage[normalem]{ulem} 
\usepackage{cancel}
\usepackage{verbatim}
\usepackage{pifont}
\usepackage{rotating}
\usepackage{subfigure}
\usepackage{multirow}
\usepackage{changepage}
\usepackage{mathtools}

\usepackage[colorlinks=true,linkcolor=red,citecolor=blue]{hyperref}

\newcommand{\be}{\begin{equation}}
\newcommand{\ee}{\end{equation}}
\newcommand{\bea}{\begin{eqnarray}}
\newcommand{\eea}{\end{eqnarray}}

\begin{document}
\setcounter{page}{1}

\title{Proton emission half-lives and shape coexistence for $71 \leq Z \leq 83$ odd-$Z$ nuclei}
\author{Yongbeom \surname{Choi}}
\email{1991.yb.choi@gmail.com}
\affiliation{School of Physics, Peng Huanwu Collaborative Center for Research and Education, and International Research Center for Big-Bang Cosmology and Element Genesis,
Beihang University, (37 Xueyuan Road, Haidian District,) Beijing 100191, People's Republic of China}
\author{Chang-Hwan \surname{Lee}}
\email{clee@pusan.ac.kr}
\affiliation{Department of Physics and Center for Innovative Physicist Education and Research, Pusan National University, Busan 46241, Korea}
\author{Youngman \surname{Kim}}
\email{ykim@ibs.re.kr}
\affiliation{ Center for Exotic Nuclear Studies, Institute for Basic Science, Daejeon 34126, Korea }

\date[]{}

\begin{abstract}
One-proton emission is a direct probe of nuclear structure near the proton drip line and plays a critical role in understanding exotic decay modes and nucleosynthesis processes. 
In this study, we investigate the half-lives of one-proton emitters for $71 \leq Z \leq 83$ odd-$Z$ nuclei by employing the WKB approximation with nuclear potentials obtained from the deformed relativistic Hartree-Bogoliubov theory in continuum (DRHBc) and, for comparison, the relativistic continuum Hartree-Bogoliubov theory (RCHB). 
We first compare the calculated half-lives with available experimental data.
The inclusion of quadrupole deformation via the DRHBc hardly contributes to improving the predictions of half-lives for the deformed nuclei.
We find that all the studied nuclei exhibit ground states with $|\beta_{2,{\rm DRHBc}}| < 0.15$, and within this limited deformation range the spectroscopic factor provides the dominant contribution to the half-life, compared to the decay width.
In particular, for nuclei exhibiting shape coexistence in DRHBc, such as $^{170}$Au, where the half-life varies significantly with the quadrupole deformation through its effect on the spectroscopic factor, we expect shape coexistence to exert a substantial influence on the variation of half-lives.
Finally, we discuss the half-lives in consideration of shape coexistence.
Our results indicate that the calculated half-life is governed not by the total-energy difference between coexisting minima but rather by the spectroscopic factor influenced by the deformation.
\end{abstract}

\maketitle


\section{Introduction}

One-proton emission is one of the nuclear decay processes that predominantly occurs in proton-rich nuclei lying near the proton drip line. 
This phenomenon provides a unique opportunity to explore the properties of exotic nuclei far from stability, shedding light on fundamental aspects of nuclear structure, such as shell structures and the continuum effects. 
The study of one-proton emitters has attracted significant interest in recent years, not only for its implications in nuclear physics but also for its relevance in astrophysical nucleosynthesis processes, particularly in the path of the rapid proton-capture process. 
In particular, a detailed understanding of proton emission half-lives is essential for interpreting experimental data, testing theoretical models, and advancing our knowledge of the underlying mechanisms inducing nuclear decay.

Since the first observation of one-proton emission from $^{53m}$Co~\cite{Jackson:1970wid, Cerny:1970zvr, Sarmiento:2023wwu}, over 60 one-proton emitters have been identified and approximately 50 half-lives have been measured~\cite{Sarmiento:2023wwu, BLANK2008403, GHCPDD}.
To understand one-proton emitters and predict their half-lives, theoretical models have played a crucial role.
In the theoretical approaches, the core of calculating the half-lives is to determine the nuclear potential between proton and daughter nucleus and estimate the tunneling probability~\cite{Janecke1965emission, Qi:2012pf, Lim:2016gwi, Deng:2019xqy, Delion:2021svs, Zou2022favored}.
The WKB (Wentzel-Kramers-Brillouin) approximation has been frequently employed to estimate the half-lives~\cite{Lim:2016gwi, Alavi:2018cuk, Sahoo:2019ytn, Zou2022favored, Xiao:2023uld, Lu:2024cko}, offering a relatively simple yet effective prediction of the half-lives of the one-proton emitters.
Using more sophisticated potentials provides a more detailed understanding of this decay process.

An additional nuclear property of particular relevance could be shape coexistence, namely the presence of two or more competing minima in the potential energy surface within a narrow energy range. Shape coexistence has been predicted in many studies~\cite{Morinaga:1956zza, Heyde:2011pgw, Garrett:2021kfb, Yang:2023dql, Bonatsos:2023xyk}. In the case of one-proton emission, shape coexistence can alter the single-particle structure of the daughter nucleus and thereby modify the spectroscopic factor, which directly enters the half-life calculation within the WKB framework. Since the half-life depends on both the tunneling probability and the spectroscopic factor, the coexistence of multiple shapes introduces additional complexity into theoretical predictions. A systematic study of this effect in proton-rich nuclei is therefore essential for understanding the interplay between nuclear deformation, shell evolution, and quantum tunneling at the limits of stability.

One of the advanced potentials can be obtained from the covariant density functional theory in continuum, such as the relativistic continuum Hartree-Bogoliubov theory (RCHB)~\cite{Zhou:2003jv, Xia:2017zka} and the deformed relativistic Hartree-Bogoliubov theory in continuum (DRHBc)~\cite{Zhou:2009sp, Li:2012gv, Li:2012xaa} with PC-PK1~\cite{Zhao:2010hi}.
The root-mean-square deviation of the predicted nuclear masses from experimental data is 8.082 MeV for RCHB with PC-PK1 and 2.562 MeV for even-$Z$ nuclei in DRHBc with the same functional~\cite{DRHBcMassTable:2024nvk}. Including the rotational correction energy reduces the root-mean-square deviation of DRHBc to 1.433 MeV.
For reference, widely used global mass models achieve significantly smaller root-mean-square deviations, such as 0.5595~MeV for FRDM2012~\cite{Moller:2015fba}, 0.561~MeV for HFB-31~\cite{PhysRevC.93.034337}, 0.375~MeV for Duflo-Zuker model~\cite{PhysRevC.52.R23}, and 0.298~MeV for WS4~\cite{WANG2014215}.
This highlights that the present DRHBc and RCHB frameworks are not optimized for global mass accuracy, but rather for a self-consistent description of continuum and deformation effects relevant to exotic nuclei.
While RCHB provides a spherical mean-field description, DRHBc extends this framework by incorporating axial deformation, which is essential for describing nuclei exhibiting shape coexistence.
These theories describe the ground states of exotic nuclei by self-consistently incorporating the pairing correlations and continuum effects. 
These features are essential to describe exotic nuclei~\cite{Meng:2005jv} and exotic phenomena such as nuclear halos~\cite{Meng:2015hta, Sun:2018ekv, Sun:2020tas}.
Especially, DRHBc, including the axial deformation degrees of freedom, shows good agreement with the available experimental data~\cite{Zhang:2020wvp, DRHBcMassTable:2022uhi, DRHBcMassTable:2022rvn, DRHBcMassTable:2024nvk, Zhang:2021ize}.
Based on these features, it has been successfully applied to investigate various exotic phenomena using DRHBc with PC-PK1.
For instance, the DRHBc has been used to understand the shifts in the neutron drip line by incorporating axial deformation degrees of freedom~\cite{In:2020asf, He:2021thz}, bubble structure, shape coexistence~\cite{Choi:2022rdj, Kim:2021skf}, and $\alpha$-decay half-lives~\cite{Choi:2023ztx}, as well as various applications~\cite{Zhang:2019qeu, Pan:2019gyo, Guo:2023ucm, Zhang:2023bqg, Zhang:2023fym, Mun:2023lfc, He:2024qgg, Zhang:2024xhy, Mun:2024ked, Mun:2024vzf} or extensions~\cite{Sun:2021nyl, Sun:2021nfb, Sun:2022gdu, Sun:2022qck, Zhang:2022yru, Pan:2024qkc}.

In this study, we employ the semiclassical WKB approximation together with nuclear potentials obtained from both the spherical RCHB framework and the axially deformed DRHBc to calculate the half-lives of one-proton emitters. 
The structure of this paper is organized as follows. 
In section \ref{Theoretical_framework}, we briefly introduce the relativistic Hartree-Bogoliubov theory in continuum and the WKB approximation adopted in this work. 
In Sec.~\ref{Results_section}, we present and discuss the calculated half-lives in comparison with experimental data, with a focus on the impact of deformation and shape coexistence using DRHBc. 
Finally, in section~\ref{summary_section}, we summarize our main findings and perspectives for future improvements.


\section{Theoretical framework} \label{Theoretical_framework}

\subsection{Relativistic Hartree-Bogoliubov Theory in Continuum}
Proton emission occurs predominantly in proton-rich exotic nuclei near the proton drip line. 
The proton Fermi surface of these nuclei becomes positive.
Therefore, it is essential to consider pairing correlations and continuum effects when studying proton emission.
In this section, we briefly introduce how RCHB and DRHBc incorporate pairing correlations and continuum effects.

The relativistic Hartree-Bogoliubov equation, which incorporates the mean-field and pairing field simultaneously, is obtained by using the variational method and the Bogoliubov transformation~\cite{Kucharek:1991arbi},
\begin{gather}
     \begin{pmatrix}
          h_{\rm D} - \lambda_\tau & \Delta \\ - \Delta^* & -h^*_{\rm D} + \lambda_\tau
     \end{pmatrix}
     \begin{pmatrix}
         U_k \\ V_k
     \end{pmatrix}
     = E_k \,
     \begin{pmatrix}
         U_k \\ V_k
     \end{pmatrix}. \label{HF_eq}
\end{gather}
In the above equation, the $h_{\rm D}$, $\lambda_\tau$, $\Delta$, $U_k$, $V_k$, $E_k$ denote the Dirac Hamiltonian for the nucleon, the Fermi surface of a nucleon, the pairing potential, the quasiparticle wave functions, and the energy of a quasiparticle state $k$, respectively.
In the present framework, particle number is conserved only through the Lagrange multiplier $\lambda_\tau$.
The Dirac Hamiltonian for the nucleon in coordinate space reads~\cite{A_textbook_of_Jie_Meng}
\be
    h_{\rm D} = \bm{\alpha} \cdot \bm{p} \, + \, \beta \left(M_{\rm n}+S_{\rm n}(\bm{r})\right) \, + \, V_{\rm n}(\bm{r}),
\ee
where $S_{\rm n}(\bm{r})$ and $V_{\rm n}(\bm{r})$ are the scalar and vector potential, respectively.
The pairing potential for the particle-particle channel reads 
\be \label{pairing_potential1}
\Delta(\bm{r}_1,\bm{r}_2) =V^{\rm pp} (\bm{r}_1,\bm{r}_2) \kappa (\bm{r}_1,\bm{r}_2),
\ee
where $\kappa=V^{\ast}U^T$ is the pairing tensor~\cite{A_textbook_of_Peter_Ring}.
In both theories, the density-dependent zero-range pairing interaction is used.
The interaction is expressed as 
\bea
V^{\rm pp} (\bm{r}_1,\bm{r}_2) = \frac{V_0}{2} \left(1 - P^{\sigma} \right)
                             \delta (\bm{r}_1 - \bm{r}_2 )
                             \left(1 - \frac{\rho(\bm{r}_1)}{\rho_{\rm sat}} \right),
\eea
where $V_0$, $(1 - P^{\sigma})/2$, and $\rho_{\rm sat}$ mean the pairing strength, the projector into the spin $S=0$ component, and the saturation density, respectively. 
The various densities $\rho_{\rm S}(\bm{r})$, $\rho_{\rm V}(\bm{r})$, and $\rho_{\rm TV}(\bm{r})$ are obtained as
\bea
\rho_{\rm S}(\bm{r}) = \sum_{k>0} \, \bar{V_k}(\bm{r}) V_k(\bm{r}), \nonumber \\
\rho_{\rm V}(\bm{r}) = \sum_{k>0} \, V^\dag_k (\bm{r}) V_k(\bm{r}), \\
\rho_{\rm TV}(\bm{r}) = \sum_{k>0} \, V^\dag_k (\bm{r}) \tau_3  V_k(\bm{r}), \nonumber 
\eea
where the quasiparticle state $k$ is summed only in the Fermi sea because of the no-sea approximation. 
The scalar potential $S_{\rm n}(\bm{r})$ and vector potential $V_{\rm n}(\bm{r})$, functions of densities, are calculated as 
\begin{equation}
\begin{aligned}
   S_{\rm n}(\bm{r}) &= \alpha_{\rm S} \rho_{\rm S} + \beta_{\rm S} \rho^2_{\rm S}
               + \gamma_{\rm S} \rho^3_{\rm S} + \delta_{\rm S} \Delta\rho_{\rm S}, \\
   V_{\rm n}(\bm{r}) &= \alpha_{\rm V} \rho_{\rm V} + \gamma_{\rm V} \rho^3_{\rm V}
             + \delta_{\rm V} \Delta \rho_{\rm V}
             + e A_0  +  \alpha_{\rm TV} \tau_3 \rho_{\rm TV}
             + \delta_{\rm TV} \tau_3 \Delta \rho_{\rm TV}.
\end{aligned}
\label{potentials}
\end{equation}
where $\alpha_{\rm S}$, $\beta_{\rm S}$, $\gamma_{\rm S}$, $\delta_{\rm S}$, $\alpha_{\rm V}$, $\gamma_{\rm V}$, $\delta_{\rm V}$, $\alpha_{\rm TV}$, and $\delta_{\rm TV}$ are the parameters for the nuclear covariant energy density functional with a point-coupling interaction. The values and the saturation density $\rho_{\rm sat}$ are taken from PC-PK1~\cite{Zhao:2010hi}. 
In the axially deformed DRHBc framework, both the densities and mean-field potentials are expressed via a multipole expansion in terms of Legendre polynomials~\cite{Price:1987sf}
\be
f(\bm{r}) = \sum_{\lambda} \, f_{\lambda} (r) P_{\lambda} (\cos \theta),
            \,\,\, \lambda = 0, \, 2, \, 4, \, \cdots.
\ee
In the DRHBc, Dirac equations are solved by expanding the wave functions in the Dirac Woods-Saxon basis. The spatial reflection symmetry is preserved by including even-order components in the Legendre polynomials.
The pairing strength is set to $V_0 = -325~{\rm MeV ~ fm^3}$, the angular momentum cutoff is $J_{\rm max}=23/2~\hbar$, and the Legendre expansion is truncated at $\lambda_{\rm max}= 8$, following Ref.~\cite{DRHBcMassTable:2024nvk}.
In RCHB, Dirac equations are solved using the shooting method in the coordinate space. The pairing strength is set to $V_0 = -342.5~{\rm MeV ~ fm^3}$, and the angular momentum cutoff is $J_{\rm max}=19/2~\hbar$, following~\cite{Xia:2017zka}.

\subsection{WKB approximation for one-proton emission\label{WKB_equations}}

\begin{figure}[h]
\begin{center}
\includegraphics[width=0.25\columnwidth]{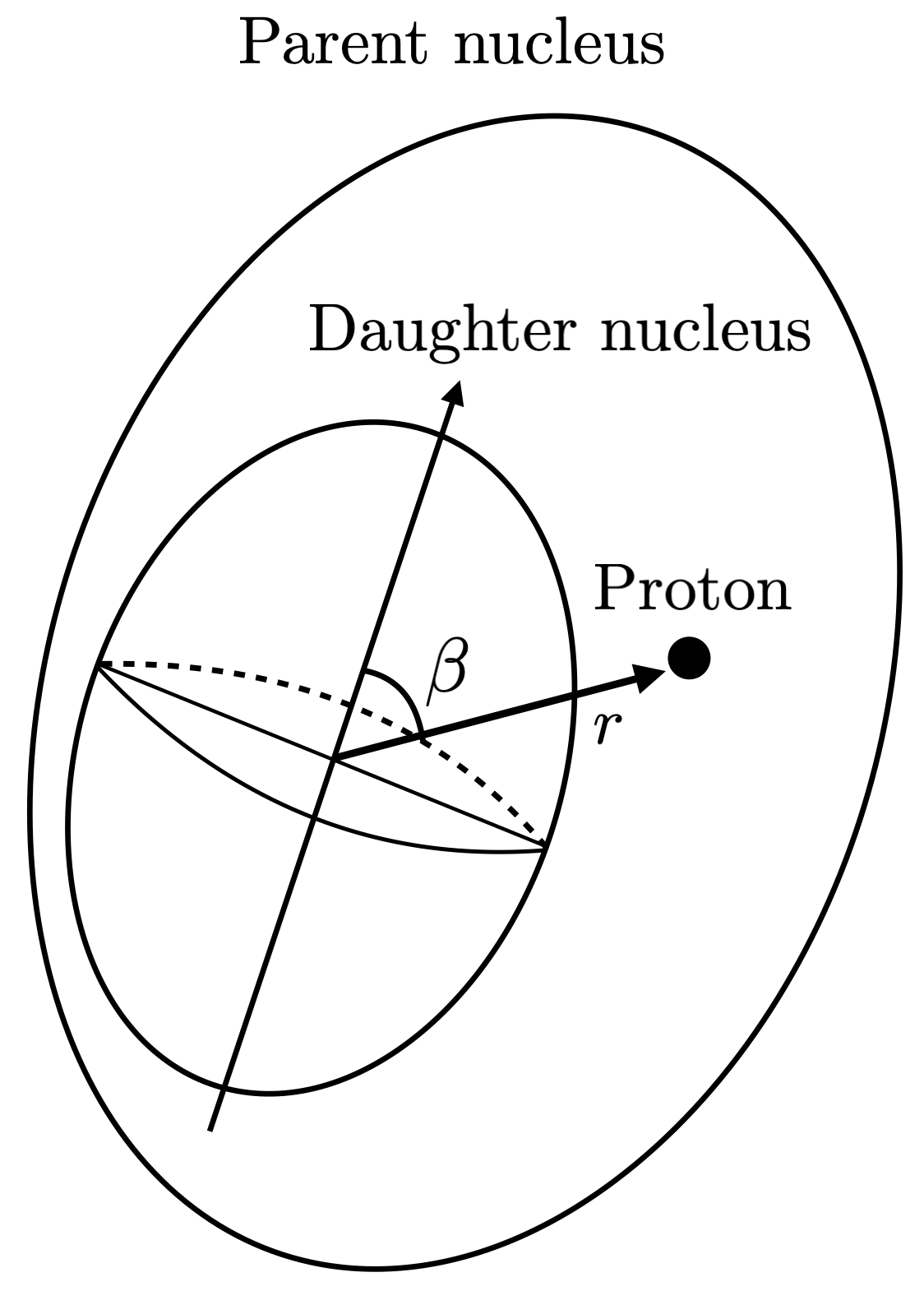}
\caption{A schematic picture to describe the orientation angle $\beta$ between the symmetry axis of the daughter nucleus and  the proton.} \label{Schematic_plot_btw_Dn_P} 
\end{center}
\end{figure}

In order to calculate the half-lives of one-proton emitters, we adopt the semiclassical WKB approximation.
In this framework, the half-lives and decay widths are calculated as 
\bea
T_{1/2} &=& \frac{\hbar \ln 2}{S_{\rm p} \Gamma} \label{T_halp_with_WKB} \label{hl_eq}, \\
\Gamma &=& N_{\rm f} \frac{\hbar^2}{4\mu} P_{\rm total}, \label{eq_gamma}
\eea
where $S_{\rm p}$ is the spectroscopic factor. 
It corresponds to the probability that adding a proton to the daughter nucleus reproduces the ground state of the parent nucleus.
This factor is typically estimated using the nonoccupation probability $u^2$ of the corresponding orbital in the daughter nucleus~\cite{Lim:2016gwi,Xiao:2023uld}.
However, DRHBc accounts for axial deformation, leading to the splitting of spherical single-particle orbitals.
In this study, we estimate $S_{\rm p}$ for the specific spherical orbital $(nlj)$ that corresponds to the experimentally assigned angular momentum $l$ of the emitted proton by expanding the DRHBc wave function $| \Psi \rangle$ in the spherical basis and evaluating the corresponding occupation probability using the following formula~\cite{Sun:2021nyl}:
\be
S_{\rm p} = 1 - \langle \Psi | \hat{N}_{nlj} | \Psi \rangle = 1- \langle \Psi | \sum_m c_{njlm}^\dagger c_{njlm} | \Psi \rangle \label{S_p_DRHBc}. 
\ee

For axially deformed nuclei, the normalization factor $N_{\rm f}$ and penetration probability $P_{\rm total}$ are calculated as
\bea
N_{\rm f} &=& \frac{1}{2} \int_{0}^{\pi} N_{\rm f}(\beta) \sin\beta d\beta,\\
P_{\rm total} &=& \frac{1}{2} \int_{0}^{\pi} P_{\rm total}(\beta) \sin \beta d\beta, \label{Gamma_halp_with_WKB}
\eea
where $\beta$ is the orientation angle between the proton and the symmetry axis of the daughter nucleus (see Fig.~\ref{Schematic_plot_btw_Dn_P}).
The normalization factor $N_{\rm f}$~\cite{Gurvitz:1986uv} and penetration probability $P_{\rm total}$ for the orientation angle $\beta$ are given by
\bea
N_{\rm f}(\beta) &=& \bigg[ \int_{r_1(\beta)}^{r_2(\beta)} \frac{d{r}'}{k(r',\beta)} \cos^2 \bigg( \int_{r_1(\beta)}^{r'} d{r}'' k(r'',\beta) - \frac{\pi}{4} \bigg) \bigg]^{-1}, \\
P_{\rm total}(\beta) &=& \exp \bigg[ -2 \int_{r_2(\beta)}^{r_3(\beta)} k(r',\beta) d{r}' \bigg].
\eea

In the above equations, $N_{\rm f}(\beta)$ and $P_{\rm total}(\beta)$ are functions of the wave number $k(r,\beta)$ calculated as
\be
k(r,\beta) = \sqrt{\frac{2\mu}{\hbar^2} |Q-V(r,\beta)| },
\ee
where $Q$, $V(r,\beta)$, and $\mu$ are the $Q$ value, the potential barrier, and the reduced mass calculated from the masses of the daughter nucleus and the proton, respectively.
The quantities $r_1(\beta)$, $r_2(\beta)$, and $r_3(\beta)$ ($r_1 < r_2 < r_3$) are the classical turning points determined by $V(r,\beta)=Q$.
The total potential barrier is composed of Coulomb, nuclear, and centrifugal contributions, and can be expressed as
\be
V(r,\beta) = V_{\rm C}(r,\beta) + V_{\rm N}(r,\beta) + \frac{\hbar^2}{2\mu} \frac{l(l+1)}{r^2}, \label{total_potential}
\ee
where $V_C(r,\beta)$ and $V_N = V_{\rm n}(r,\beta) + S_{\rm n}(r,\beta)$ are obtained from the daughter nucleus calculated from the relativistic Hartree-Bogoliubov theories in continuum.
The same experimentally assigned angular momentum $l$ is consistently adopted in both the evaluation of the spectroscopic factor (Eq.~\ref{S_p_DRHBc}) and the centrifugal barrier term in the WKB calculation (Eq.~\ref{total_potential}). 
In this study, the spectroscopic factor and the decay width are evaluated for the same specific partial wave.
A fully consistent treatment would require angular-momentum projection and an explicit decomposition into partial decay widths, which is beyond the scope of the present work.

\section{Results\label{Results_section}}
In this section, we present the calculated proton emission half-lives of $71 \leq Z \leq 83$ odd-$Z$ nuclei within the WKB approximation framework.
For each nucleus, two half-lives are obtained using nuclear potentials derived from DRHBc and RCHB. 
We first discuss the results for $^{150,151}$Lu, $^{155}$Ta, $^{159}$Re, $^{164-167}$Ir, and $^{185}$Bi in Sec.~\ref{Results_subsection1}.
In cases of shape coexistence, the DRHBc framework may provide multiple half-lives for a single nucleus.
We next analyze these cases for $^{156,157}$Ta, $^{160,161}$Re, $^{170,171}$Au, and $^{176,177}$Tl in Sec.~\ref{Results_subsection2}
To calculate the half-lives, we take the $Q$ value and the orbital of the emitted proton from Ref.~\cite{BLANK2008403}.
It is worth noting that the calculated half-life depends exponentially on the $Q$ value. 
For example, in the case of $^{177m}$Tl, which has the largest $Q$ value among the studied nuclei ($Q = 1.984$ MeV), a variation of only 0.1 MeV in $Q$ leads to a change of the half-life by approximately a factor of 4.
This illustrates that even small uncertainties in experimental or theoretical $Q$ values can induce sizable deviations in the predicted half-lives.
The nuclear potential of the isomeric state is taken to be the same as that of the ground state.
This is because the potential itself remains nearly unchanged and the large differences in half-lives are mainly attributed to the angular momentum of the emitted proton and the corresponding $Q$ value~\cite{Lim:2016gwi}.
The calculated half-lives are shown in Figs.~\ref{fig2}-\ref{fig6} and listed in Tables~\ref{tab1} and \ref{tab2}.

\subsection{Proton emission half-lives\label{Results_subsection1}}
\begin{figure}[h]
\begin{center}
\includegraphics[width=0.45\columnwidth]{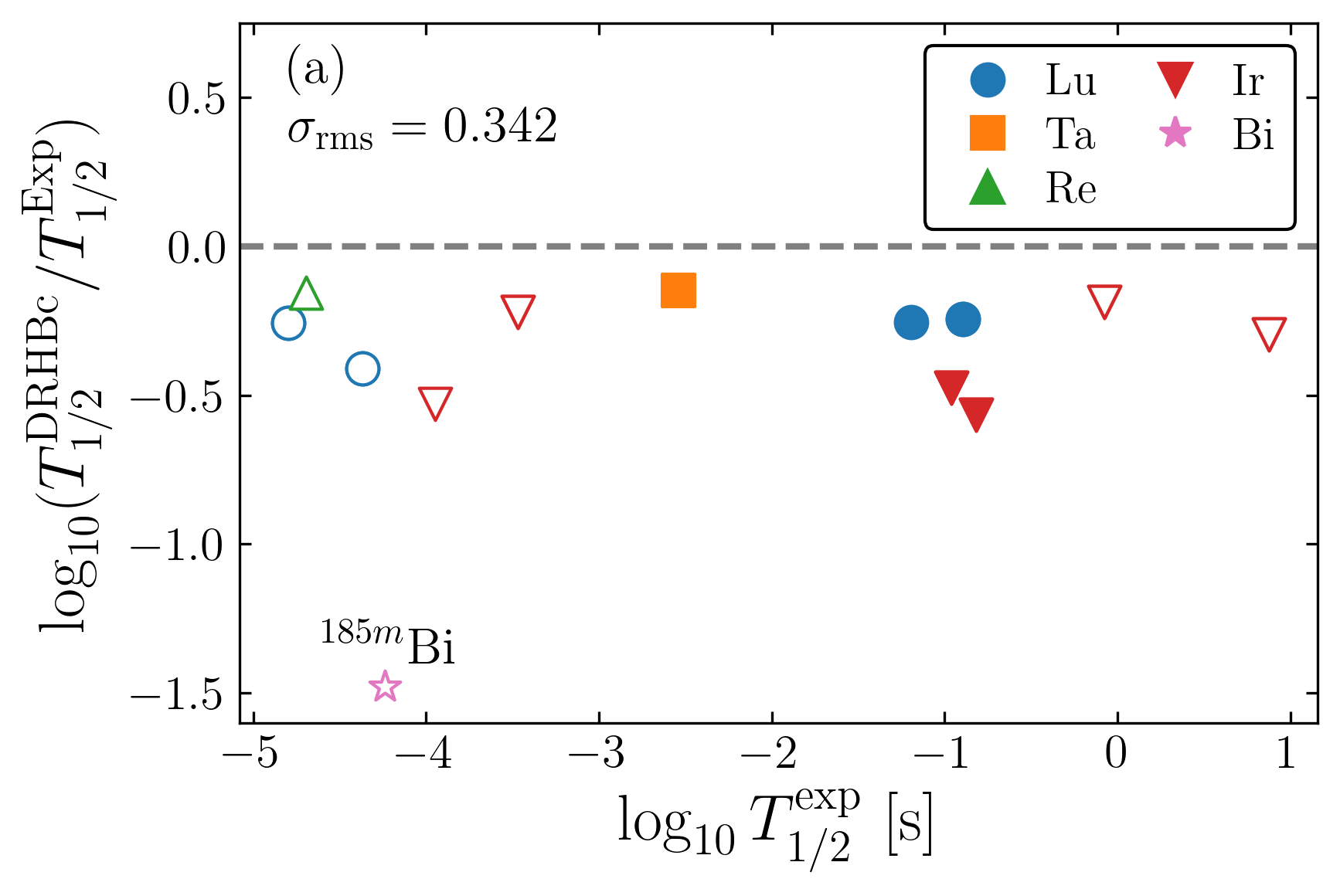}
\includegraphics[width=0.45\columnwidth]{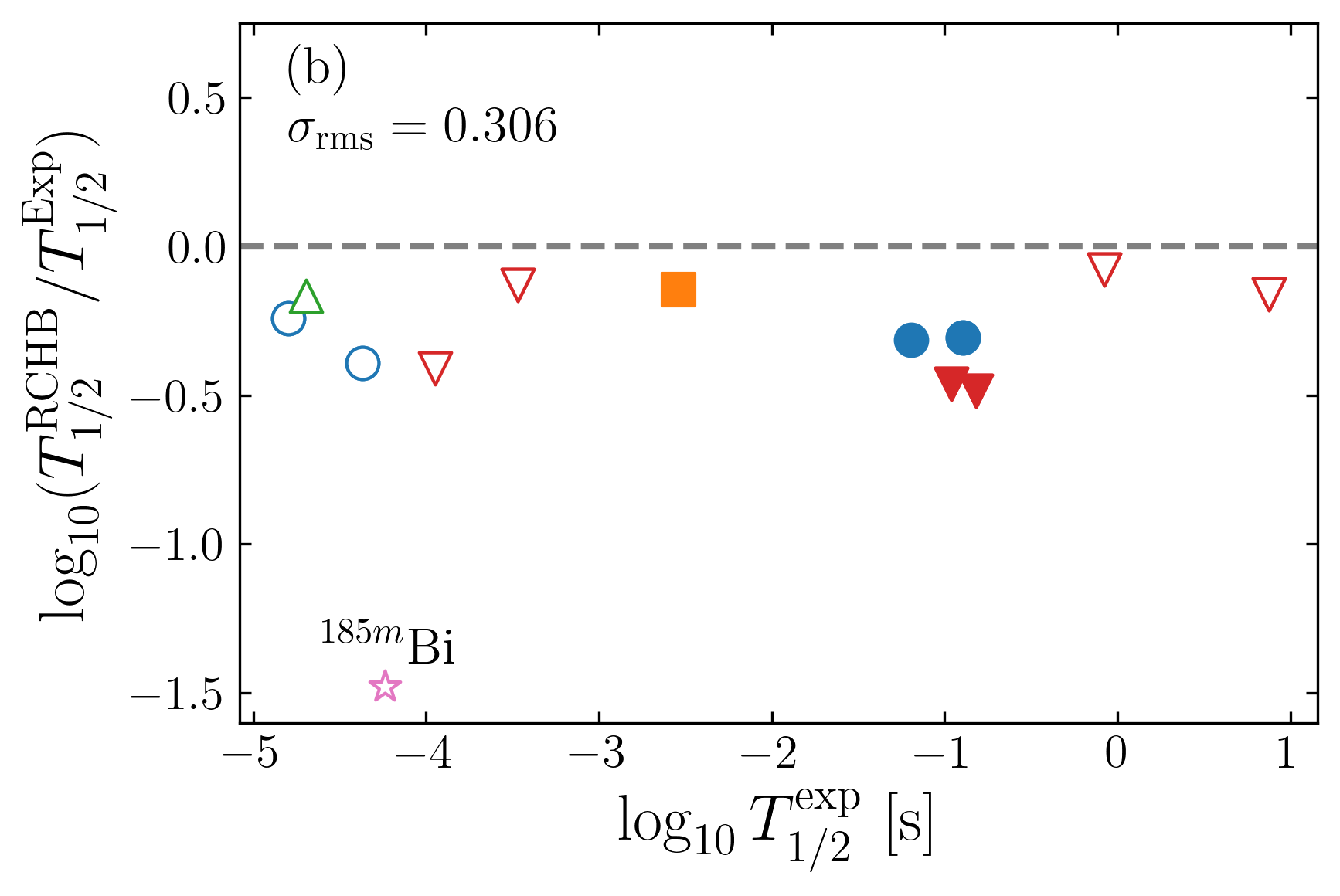}
\includegraphics[width=0.45\columnwidth]{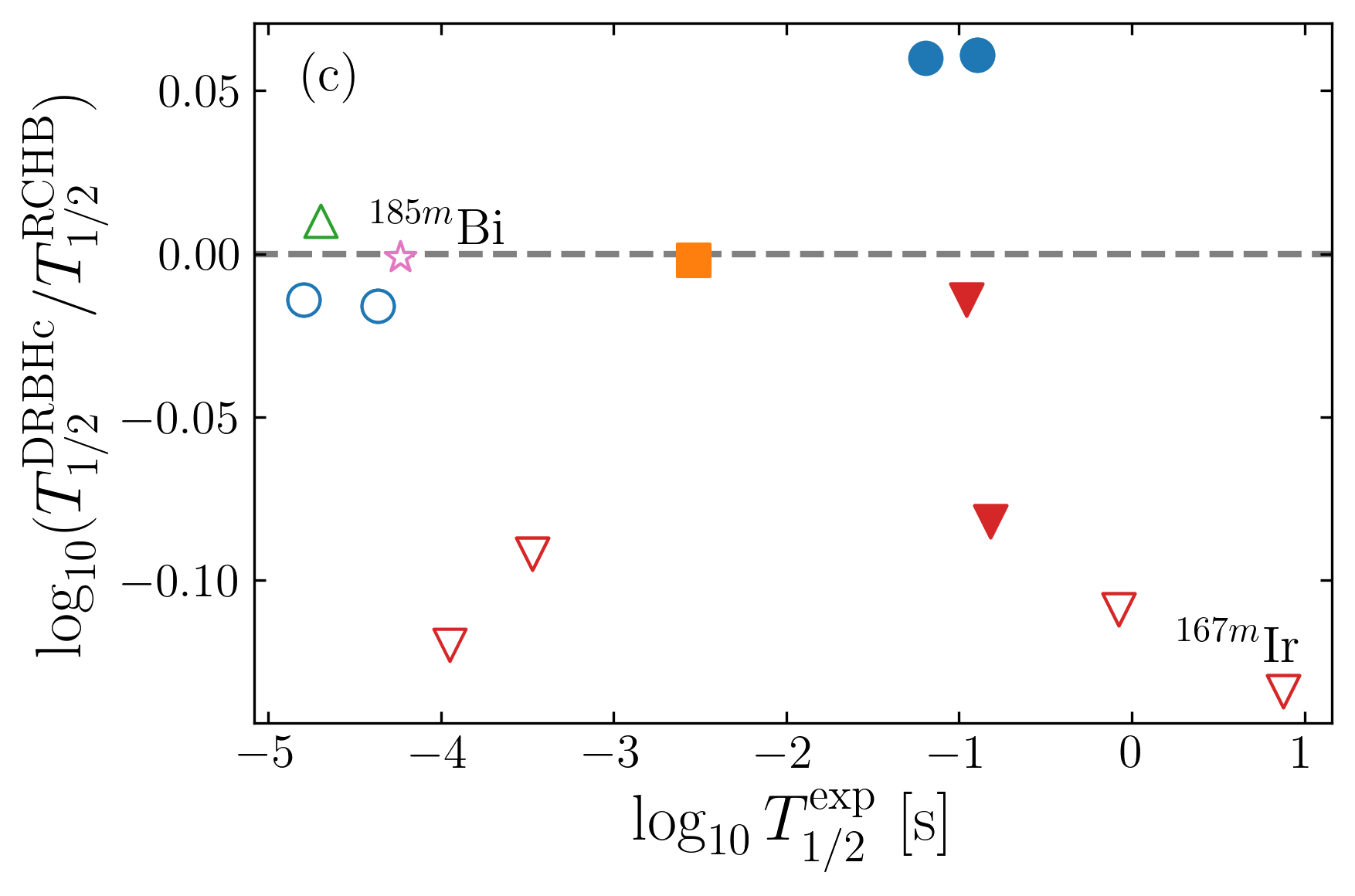}
\caption{Comparison of the calculated one-proton emission half-lives with experimental data~\cite{BLANK2008403}.
(a) The deviations of the half-lives calculated by DRHBc from the experimental values.
(b) The same as (a) but for the results obtained using RCHB.
(c) The ratio between the half-lives calculated by DRHBc and RCHB.
The empty markers denote the isomeric states.
The root-mean-square deviations $\sigma_{\rm rms}$ are indicated in panels (a) and (b).
} \label{fig2} 
\end{center}
\end{figure}

To quantitatively compare the calculated half-lives with the experimental data~\cite{BLANK2008403}, we introduce the root-mean-square deviation defined as
\bea
\sigma = \sqrt{\frac{1}{N} \sum^N_{i=1} \left(
\log_{10} T^{\rm cal}_{1/2, i} 
- \log_{10} T^{\rm exp}_{1/2, i} \right)^2},
\label{eq:rms_deviation}
\eea
where $T^{\rm cal}_{1/2, i}$ and $T^{\rm exp}_{1/2, i}$ are the calculated and experimental half-lives, respectively.
The nuclide $^{185m}\mathrm{Bi}$ is excluded from the calculation of $\sigma$, since the deviation from the experimental value is exceptionally large (see Fig~\ref{fig2}).

Figure~\ref{fig2}(a) and (b) display the deviations between calculated half-lives and experimental data.
In Fig.~\ref{fig2}(a), DRHBc is used to calculate the half-lives, while Fig.~\ref{fig2}(b) shows the results obtained with RCHB.
Except for $^{185m}$Bi, the half-lives calculated with both DRHBc and RCHB systematically underestimate the experimental half-lives by up to a factor of about 4 ($\simeq 10^{-0.6}$), and a similar trend is also observed in Fig.~\ref{fig5}.
This discrepancy might originate from the overestimation of the decay width.
The overall trends are similar between the ground states and the isomeric states (empty markers), without any obvious correlation between the deviations and the experimental half-lives.
In the case of $^{185m}$Bi, both DRHBc and RCHB deviate significantly from the experimental value. 
However, the predictions of DRHBc and RCHB for $^{185m}$Bi are nearly identical, as shown in Fig.~\ref{fig2}(c).
Bi ($Z$ = 83) lies just above the proton shell closure at $Z$ = 82, whereas all other emitters considered in this study have atomic numbers below 82. 
The large deviation from the experiment in the case of a Pb core coupled with one additional proton may indicate the intrinsic limitations of the mean-field description, particularly in the estimation of the spectroscopic factor. 
Within the present mean-field framework, the spectroscopic factor is obtained close to unity for such a closed-shell plus one-proton system. 
A more reliable assessment would require calculations beyond the mean-field approach. 
However, even if the spectroscopic factor was reduced by a factor of two or three, the discrepancy in the half-life would still remain close to an order of magnitude. 
This suggests that the discrepancy is unlikely to be fully resolved solely through the spectroscopic factor and may reflect an enhanced sensitivity of the decay width to the mean-field potential in such closed-shell plus one-proton systems.
For the other nuclei, the difference between DRHBc and RCHB is modest, with the largest deviation being about a factor of 1.3 for $^{167m}$Ir (see Fig.~\ref{fig2}(c)).
Note that the pairing strengths adopted in DRHBc and RCHB differ by about 5\% ($V_0 = -325$ MeV fm$^3$ and $-342.5$ MeV fm$^3$, respectively). 
This suggests that moderate variations in the pairing strength have only a limited influence on the spectroscopic factors and the calculated half-life predictions, as illustrated by the case of $^{155}$Ta, where both the parent and daughter nuclei are nearly spherical and the calculated half-lives from DRHBc and RCHB remain nearly identical.
The root-mean-square deviation of the logarithmic half-lives is estimated to be 0.342 for DRHBc and 0.306 for RCHB.
The comparison of the root-mean-square deviations indicates that deformation has a non-negligible influence on the calculated half-lives, as variations in $\beta_2$ can also affect the spectroscopic factor $S_{\rm p}$.

\begin{figure}[t]
\begin{center}
\includegraphics[width=0.45\columnwidth]{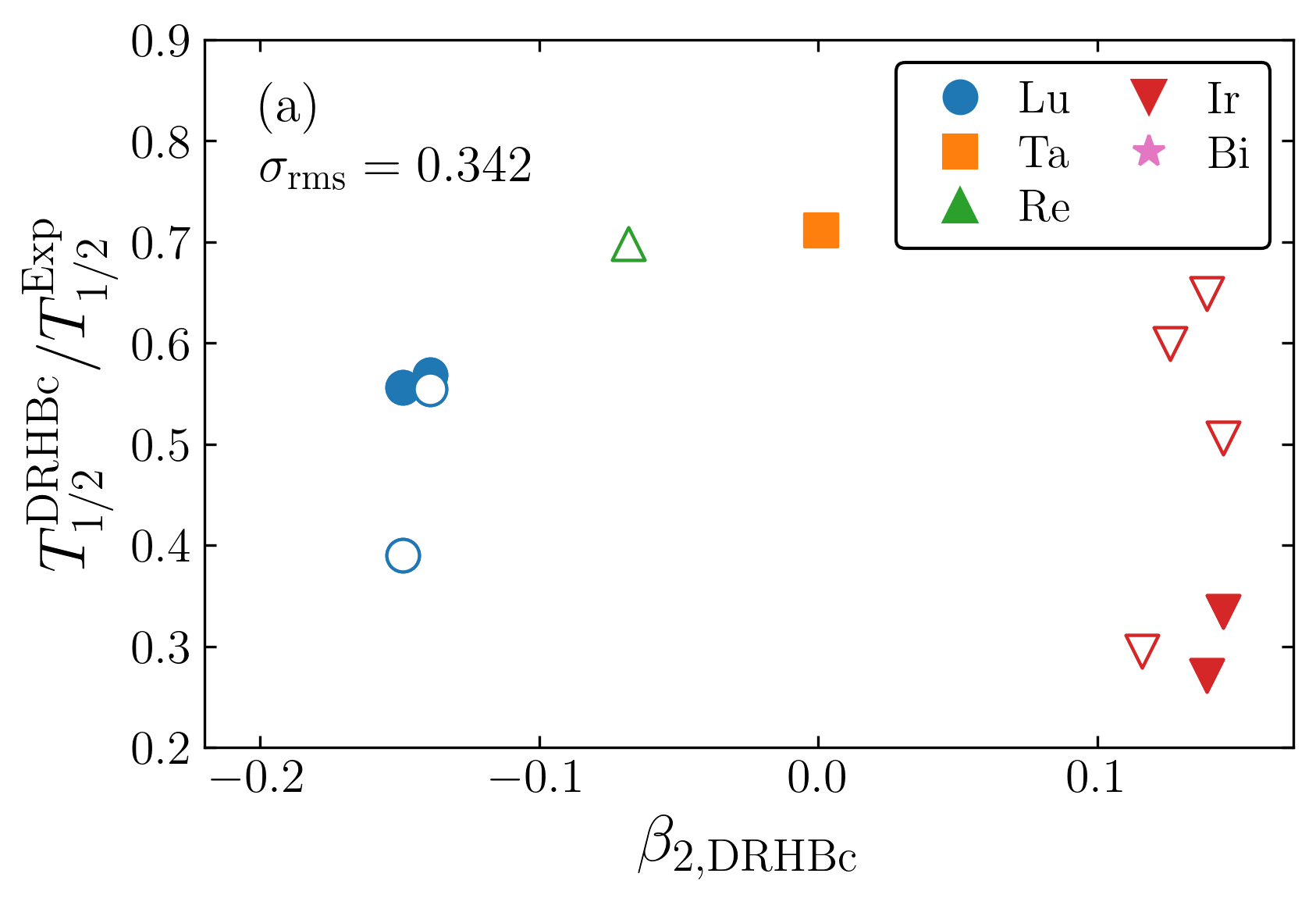}
\includegraphics[width=0.45\columnwidth]{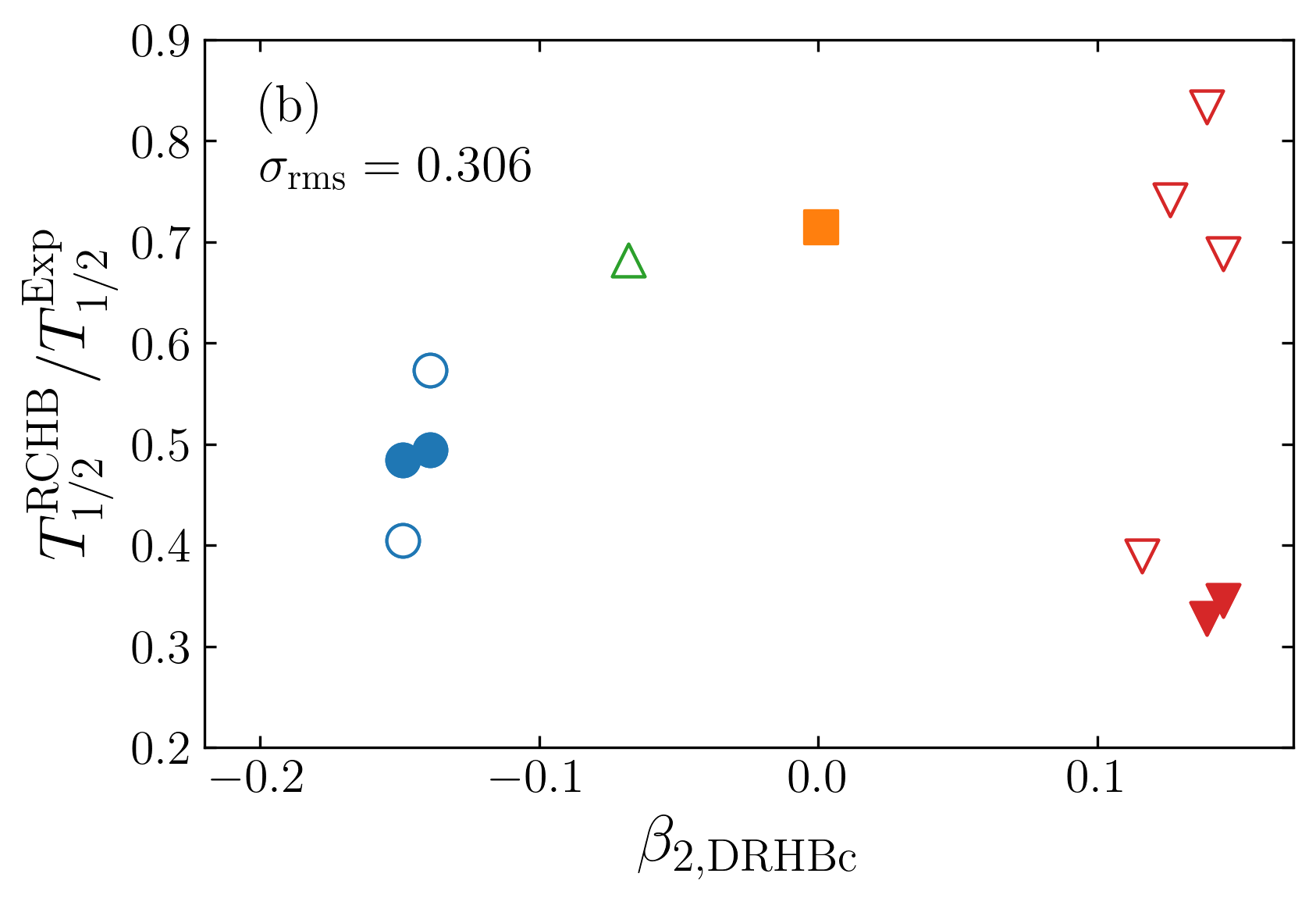}
\includegraphics[width=0.45\columnwidth]{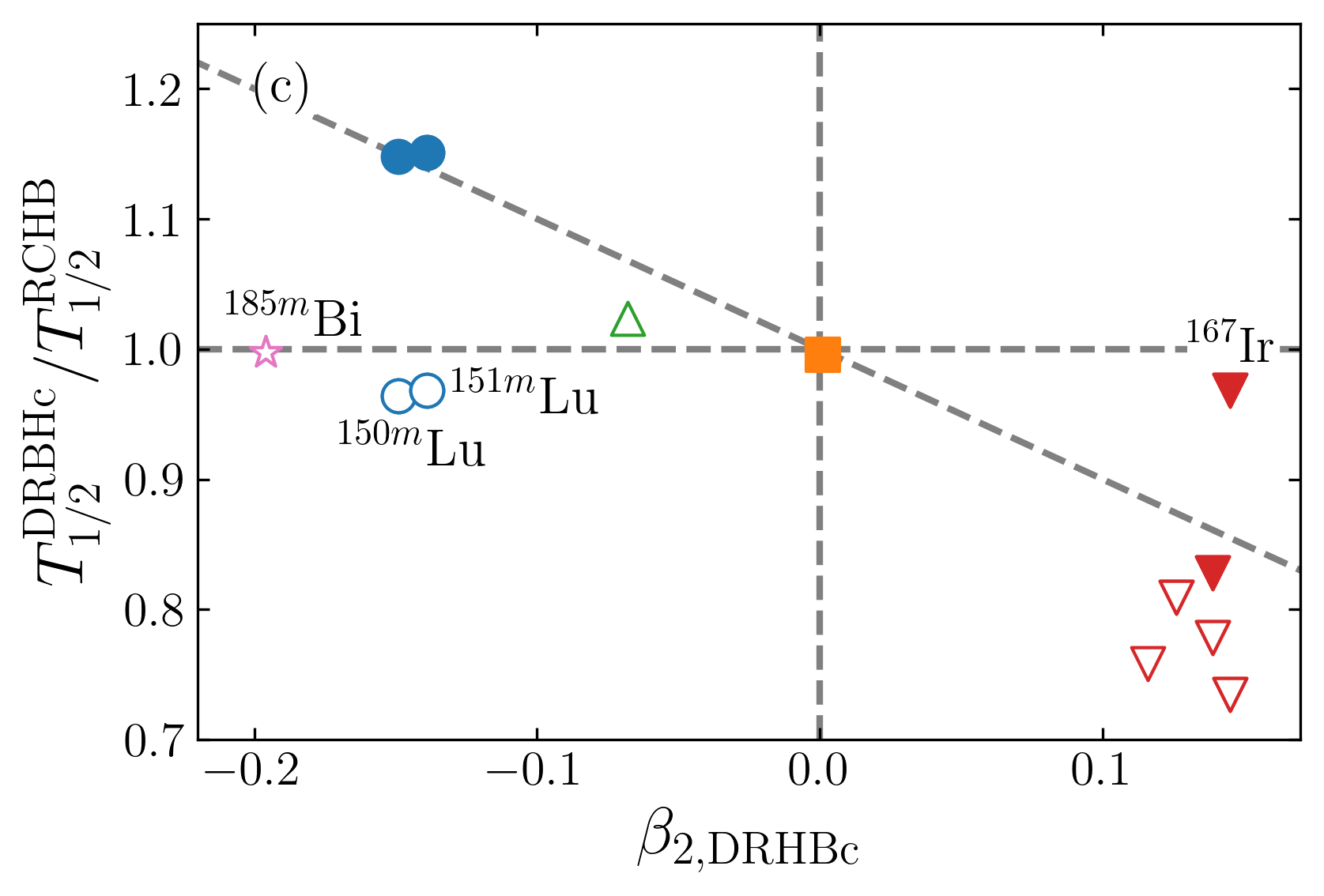}
\includegraphics[width=0.45\columnwidth]{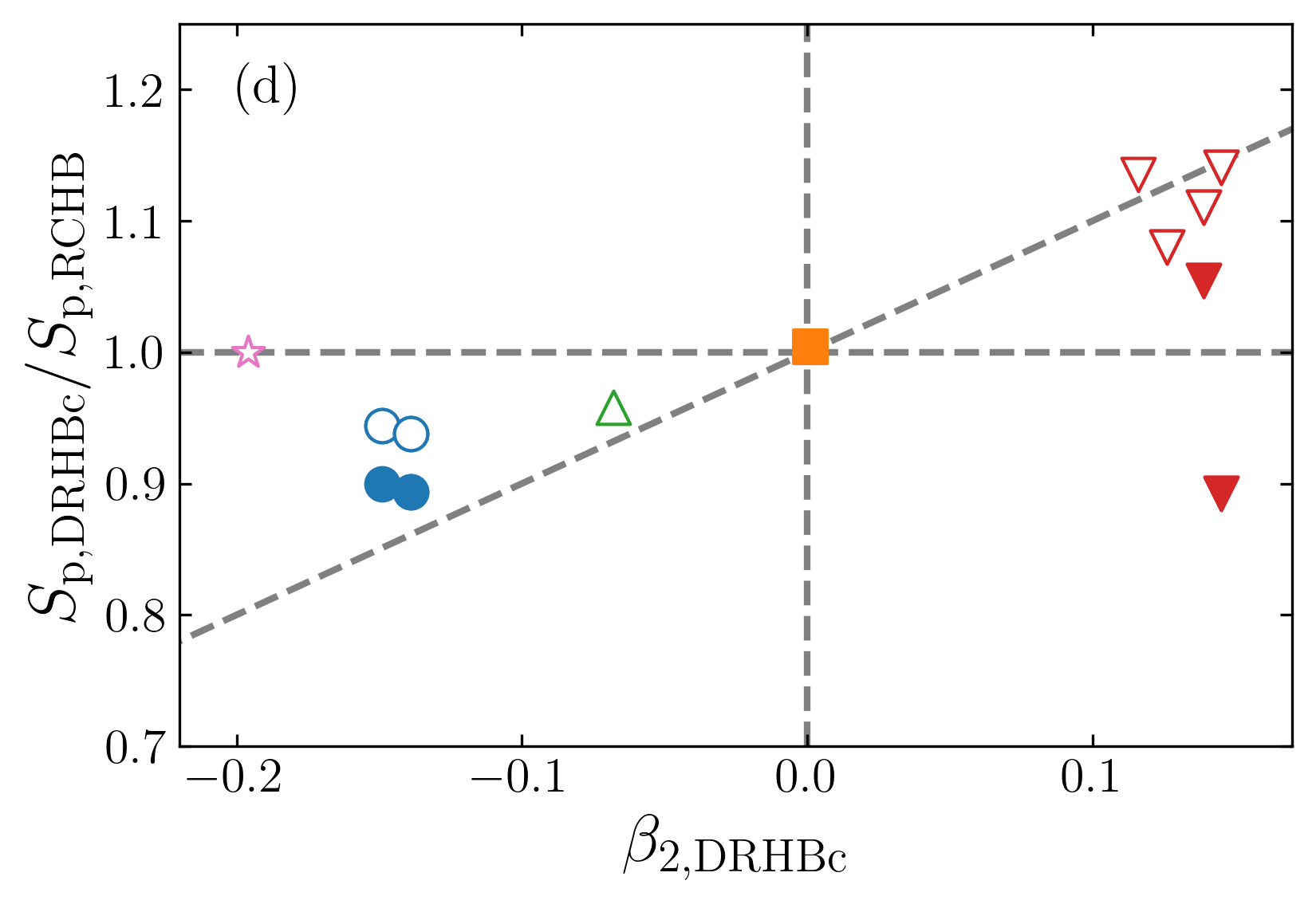}
\includegraphics[width=0.45\columnwidth]{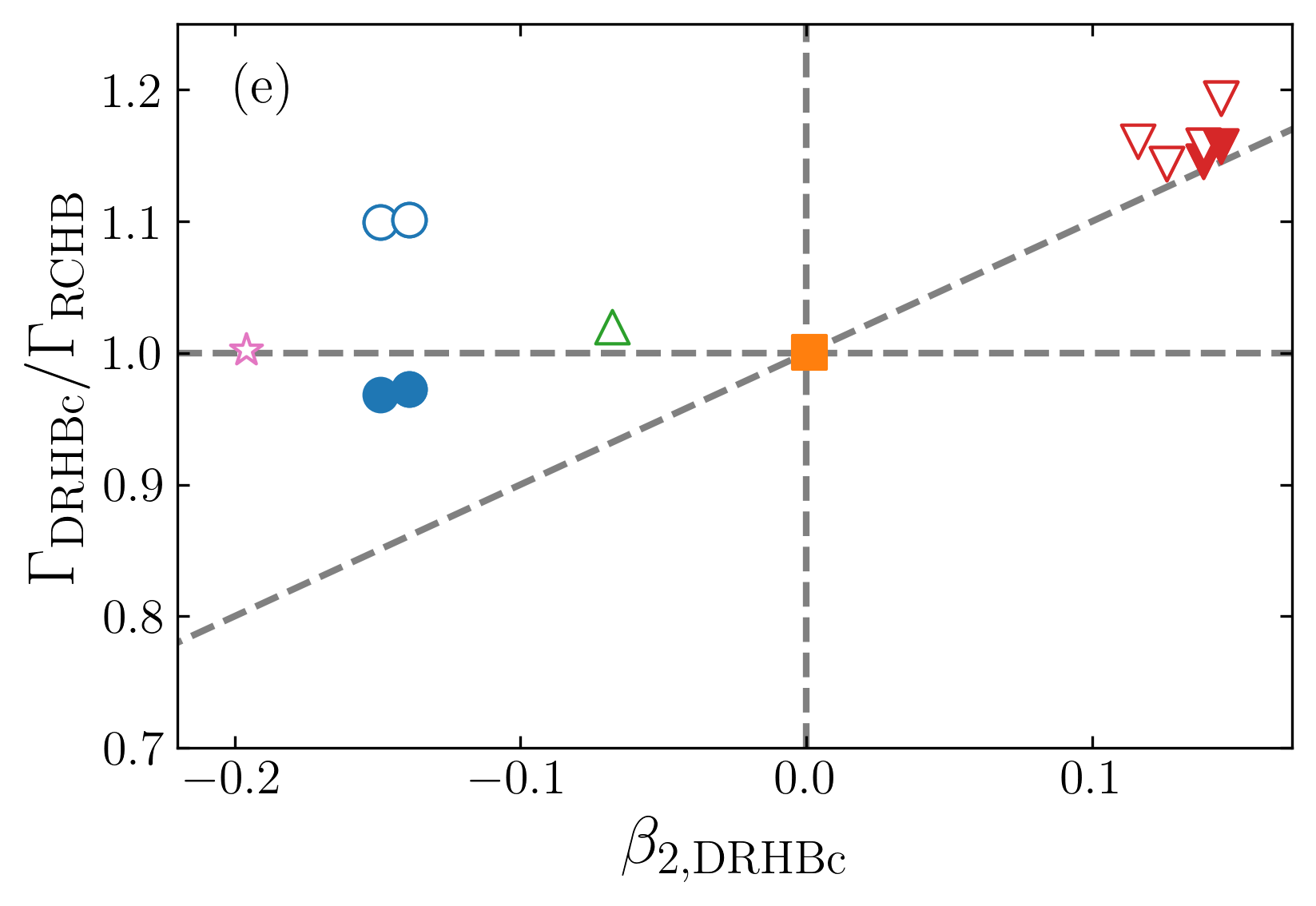}
\caption{Comparison of calculated half-lives, spectroscopic factors, and decay widths as functions of quadrupole deformation. 
(a) Comparison between half-lives using DRHBc with experimental data. (b) Comparison between half-lives using RCHB with experimental data. (c) DRHBc vs RCHB half-lives. (d) Comparison between spectroscopic factor using DRHBc with those using RCHB. (e) Comparison between decay width using DRHBc with those using RCHB. For $^{185m}$Bi, $T^{\rm DRHBc}_{1/2}/T^{\rm exp}_{1/2}$ = $T^{\rm RCHB}_{1/2}/T^{\rm exp}_{1/2}$ = 0.331.} \label{fig3} 
\end{center}
\end{figure}

In the next step, we examine the calculated values from DRHBc and RCHB as functions of the quadrupole deformation $\beta_{2,{\rm DRHBc}}$, as illustrated in Fig.~\ref{fig3}.
Figure~\ref{fig3}(a) and (b) compare the calculated half-lives with the experimental data, while Fig.~\ref{fig3}(c) shows the comparison between DRHBc and RCHB.
From Fig.~\ref{fig3}(a) and (b), it can be seen that nuclei with larger values of $|\beta_{2,{\rm DRHBc}}|$ are not necessarily better reproduced even adopting DRHBc, indicating that the agreement with experiment is independent of the quadrupole deformation.
Moreover, no clear systematic trend is observed in the deviations of half-lives as a function of $Q$ value, proton number $Z$, mass number $A$, or neutron number $N$.
As shown in Fig.~\ref{fig3}(c), when directly comparing DRHBc with RCHB, a correlation with $\beta_{2,{\rm DRHBc}}$ can be identified, except for $^{185m}$Bi, $^{150m,151m}$Lu, and $^{167}$Ir.
However, in cases such as $^{185m}$Bi, $^{150m,151m}$Lu, and $^{167}$Ir, the half-lives predicted by DRHBc and RCHB are independent of quadrupole deformations.
This behavior can be understood by considering the spectroscopic factor ratio $S_{\rm p, DRHBc}/S_{\rm p, RCHB}$ and decay width ratio $\Gamma_{\rm DRHBc}/\Gamma_{\rm RCHB}$ shown in Fig.~\ref{fig3}(d) and (e).
The half-life is inversely proportional to both the spectroscopic factor $S_{\rm p}$ and the decay width $\Gamma$ (see Eq.~\ref{hl_eq}).
For $^{150,151}$Lu and $^{167}$Ir, the variation of the spectroscopic factor plays an important role in the variation of the half-lives, while the decay width ratio remains nearly unchanged.
In the case of $^{150m,151m}$Lu, both the spectroscopic factor and the decay width contribute to determining the half-life.

\begin{figure}[t]
\begin{center}
\includegraphics[width=0.95\columnwidth]{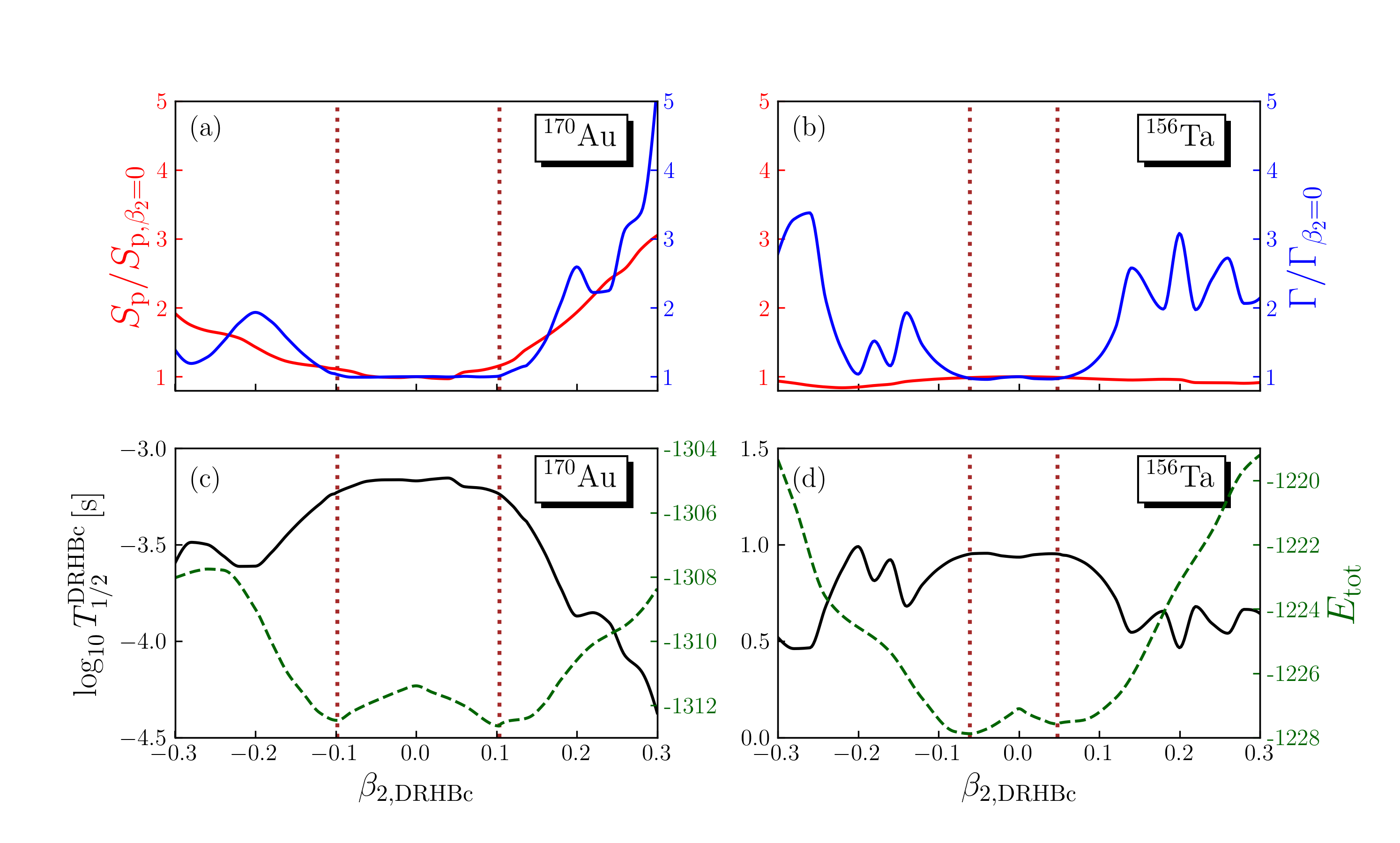}
\caption{Spectroscopic factors, decay widths, half-lives, and potential energies of $^{170}$Au and $^{156}$Ta as functions of $\beta_{2,{\rm DRHBc}}$. (a) Spectroscopic factor and decay width of $^{170}$Au. (b) Spectroscopic factor and decay width of $^{156}$Ta. (c) Half-life and potential energy of $^{170}$Au. (d) Half-life and potential energy of $^{156}$Ta. Both isotopes exhibit shape coexistence. The vertical dashed lines indicate the ground states ($|\Delta E_{\rm tot}| \le$0.3 MeV). The orbitals of emitted protons are both $1h_{11/2}$. } \label{fig4} 
\end{center}
\end{figure}

To clarify why the spectroscopic factor has a significant impact on the half-life, we investigate its dependence on the deformation together with the corresponding decay width. 
As representative examples, we consider $^{170}$Au and $^{156}$Ta.
Figure~\ref{fig4} shows the  normalized spectroscopic factors ($S_{\rm p}/S_{{\rm p}, \beta_2=0}$), and normalized decay widths ($\Gamma/\Gamma_{\beta_2=0}$), half-lives, and potential energies as functions of quadrupole deformation $\beta_{2,{\rm DRHBc}}$.
Both $^{170}$Au and $^{156}$Ta exhibit shape coexistence (with total-energy differences between the two minima of $|\Delta E_{\rm tot}| < 0.3$ MeV) in the DRHBc calculation, with two nearly degenerate ground states indicated by the vertical dashed lines. 
The values are summarized in Table~\ref{tab2}.
For $^{170m}$Au, within this range, the variation of the spectroscopic factor ($S_p/S_{p,\beta_2=0}$: 1.15 in prolate minimum and 1.11 in oblate minimum) is more pronounced than that of the decay width ($\Gamma/\Gamma_{\beta_2=0}$: 1.00 in prolate minimum and 1.04 in oblate minimum), indicating that the spectroscopic factor plays the dominant role in determining the half-life.
This trend is also confirmed in Fig.~\ref{fig6}(a) and (b).
Note that all the studied ground states lie within $|\beta_2|\lesssim0.15$.
If shape coexistence occurs in regions of larger deformation, the difference in half-lives between coexisting shapes may become even more pronounced.
As in the case of $^{156}$Ta, in regions of larger deformation the decay width may have a more significant impact on the half-life.

\begin{table*}[h!]
\centering
\caption{Calculated half-lives with DRHBc and RCHB. The Q-values, the orbital of the emitted proton, and the experimental half-lives are taken from Ref.~\cite{BLANK2008403}. For $^{185m}$Bi, the change of $\beta_2$ between parent and daughter is noticeable and the spectroscopic factor is estimated as $S_{\rm p} = 1$.} \label{tab1}
\begin{tabular}{|c|c|c|c|c|c|c|c|c|c|}
\hline
Parent & \multicolumn{2}{c|}{$\beta_{2,{\rm DRHBc}}$} & $Q$ & \multirow{2}{*}{Orbital} & \multicolumn{2}{c|}{$S_{\rm p}$} & \multicolumn{3}{c|}{$\log_{10} T_{1/2}$ [s]} \\ 
\cline{2-3} \cline{6-7} \cline{8-10}
nucleus & Parent & Daughter & [MeV] & & DRHBc & RCHB & Exp & DRHBc & RCHB \\ 
\hline
$^{\phantom{m}150}$Lu & $-$0.149 & $-$0.152  &  1.283 & $1h_{11/2}$ &  0.547 &  0.608 & $-$1.194 & $-$1.449 & $-$1.509  \\
$^{150m}$Lu & $-$0.149 & $-$0.152  &  1.306 & $2d_{3/2}$ &  0.521 &  0.552 & $-$4.367 & $-$4.776 & $-$4.760  \\
$^{\phantom{m}151}$Lu & $-$0.139 & $-$0.142  &  1.253 & $1h_{11/2}$ &  0.545 &  0.610 & $-$0.896 & $-$1.141 & $-$1.202  \\
$^{151m}$Lu & $-$0.139 & $-$0.142  &  1.332 & $2d_{3/2}$ &  0.543 &  0.579 & $-$4.796 & $-$5.052 & $-$5.038  \\
$^{\phantom{m}155}$Ta & \phantom{$-$}0.001 & \phantom{$-$}0.000  &  1.468 & $1h_{11/2}$ &  0.498 &  0.496 & $-$2.538 & $-$2.686 & $-$2.684  \\
$^{159m}$Re & $-$0.068 & $-$0.088  &  1.831 & $1h_{11/2}$ &  0.367 &  0.383 & $-$4.695 & $-$4.851 & $-$4.861  \\
$^{164m}$Ir & \phantom{$-$}0.116 & \phantom{$-$}0.137  &  1.844 & $1h_{11/2}$ &  0.285 &  0.251 & $-$3.947 & $-$4.477 & $-$4.357  \\
$^{165m}$Ir & \phantom{$-$}0.126 & \phantom{$-$}0.139  &  1.733 & $1h_{11/2}$ &  0.269 &  0.249 & $-$3.469 & $-$3.691 & $-$3.599  \\
$^{\phantom{m}166}$Ir & \phantom{$-$}0.139 & \phantom{$-$}0.150  &  1.168 & $2d_{3/2}$ &  0.406 &  0.385 & $-$0.818 & $-$1.385 & $-$1.303  \\
$^{166m}$Ir & \phantom{$-$}0.139 & \phantom{$-$}0.150  &  1.340 & $1h_{11/2}$ &  0.262 &  0.236 & $-$0.076 & $-$0.264 & $-$0.155  \\
$^{\phantom{m}167}$Ir & \phantom{$-$}0.145 & \phantom{$-$}0.161  &  1.096 & $3s_{1/2}$ &  0.606 &  0.679 & $-$0.959 & $-$1.435 & $-$1.421  \\
$^{167m}$Ir & \phantom{$-$}0.145 & \phantom{$-$}0.161  &  1.261 & $1h_{11/2}$ &  0.268 &  0.235 & \phantom{$-$}0.875 & \phantom{$-$}0.579 & \phantom{$-$}0.713  \\
$^{185m}$Bi & $-$0.196 & \phantom{$-$}0.000  &  1.624 & $4s_{1/2}$ &  1.000 &  1.000 & $-$4.237 & $-$5.717 & $-$5.716  \\
\hline
\end{tabular}
\end{table*}

\newpage

\subsection{Proton emission half-lives with shape coexistence \label{Results_subsection2}}
In addition to comparing calculated proton emission half-lives of spherical or axially deformed isotopes, we also consider the influence of nuclear shape coexistence predicted by the DRHBc framework. 
The shape coexistence is defined in this study as the presence of multiple degenerate minima with $|\Delta E_{\rm tot}| \leq 0.6$ MeV. 
Since nuclei are finite quantum many-body systems, the true ground state corresponds to a linear superposition of the wave functions associated with the degenerate minima. 
A rigorous quantum-mechanical treatment therefore requires configuration mixing, which is beyond the scope of the present study. Instead, we adopt a simple averaging procedure as a first-order approximation to estimate the sensitivity of proton-decay observables to competing nuclear shapes. In this simplified treatment, interference effects and mixing amplitudes between different configurations are not explicitly taken into account. Given the small energy difference between the two minima ($|\Delta E_{\rm tot}| \leq$ 0.6 MeV), treating them with a simple averaging procedure is expected to provide a reasonable first-order estimate of their influence on the decay observables.
For a simple two-state mixing model that describes shape coexistence, we refer to Ref.~\cite{Casten1990}.

\begin{figure}[t]
\begin{center}
\includegraphics[width=0.45\columnwidth]{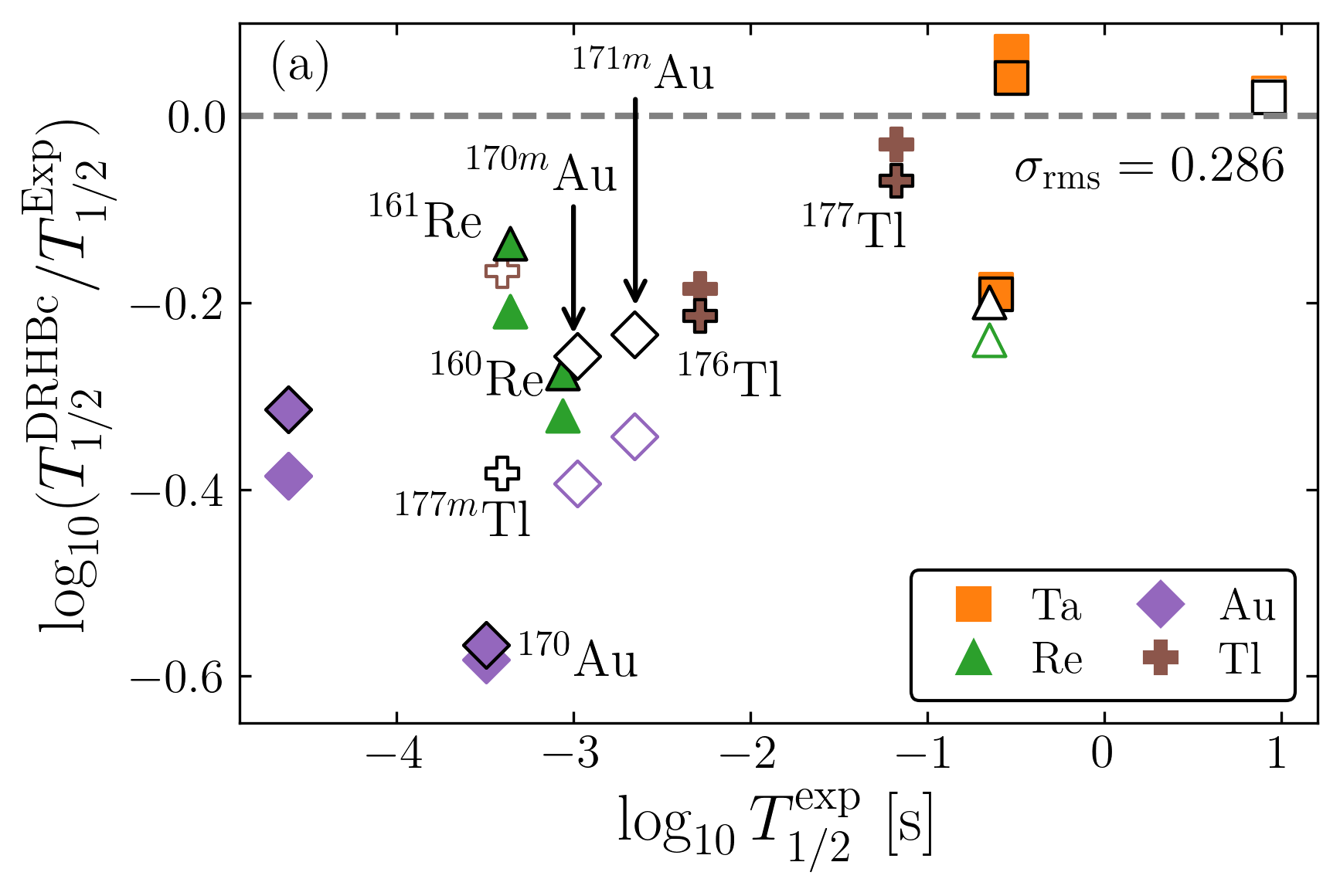}
\includegraphics[width=0.45\columnwidth]{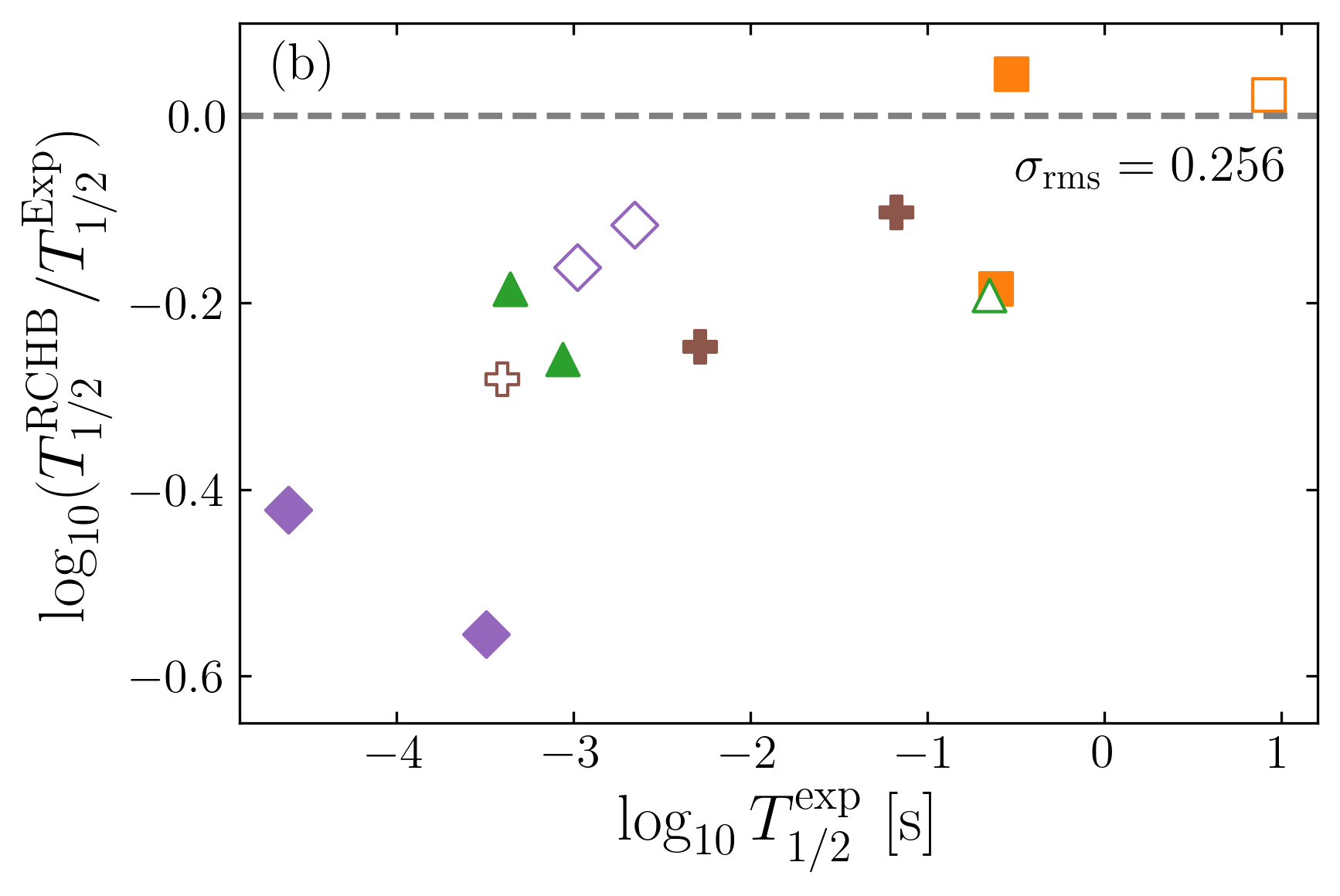}
\caption{Comparison of half-lives for nuclei with shape coexistence. Black-edged symbols denote the half-lives calculated from $E_{\rm tot} - |\Delta E_{\rm tot}|$ between the prolate and oblate configurations, whereas symbols without black edges denote those obtained from $E_{\rm tot} + |\Delta E_{\rm tot}|$. The empty markers correspond to the half-lives of the isomeric states. When the half-lives from the shape coexistence configurations are uniformly applied by averaging the logarithmic half-lives, the root-mean-square deviation is reduced to $\sigma_{\rm rms,SC}=0.272$.
} \label{fig5} 
\end{center}
\end{figure}

Before discussing the results shown in Fig.~\ref{fig5}, we point out that the shape of the daughter nucleus is taken to follow that of the parent nucleus (e.g., prolate–prolate or oblate–oblate).
This treatment is motivated by the phenomenon of shape inheritance, observed for example in $\alpha$ decay, where the deformation properties (such as prolate, oblate, or hexadecapole shapes) of a parent nucleus are partially transmitted to the resulting daughter nucleus~\cite{Li2025}.
It should be noted, however, that this treatment is a simplifying assumption, and its quantitative accuracy may depend on the parent-daughter configuration overlap.
Figure~\ref{fig5}(a) shows the comparison between calculated and experimental half-lives for nuclei exhibiting shape coexistence within the DRHBc framework, while Fig.~\ref{fig5}(b) presents the corresponding results using RCHB framework. 
It is seen that considering shape coexistence could lead to nonnegligible changes in the calculated half-lives, which may either improve or worsen the discrepancy with experiment.
For instance, for $^{177m}$Tl, the discrepancies between theory and experiment are found to be 1.47 ($\simeq 10^{|-3.568+3.402|}$) and 2.42 ($\simeq 10^{|-3.785+3.402|}$) for the two coexisting shapes, respectively. 
Since the true ground state is a linear combination of the wave functions associated with the degenerate minima, we take a simple average of these discrepancies.
Therefore, the expected discrepancy is 1.88 ($\simeq 10^{|-3.677+3.402|}$) when shape coexistence is taken into account by a simple averaging.
Similarly, for $^{170m}$Au and $^{171m}$Au, the deviations decrease from 2.48 ($\simeq10^{|-3.374+2.980|}$) and 2.20 ($\simeq10^{|-2.997+2.654|}$) to 2.12 and 1.94, respectively, when the effect of shape coexistence is considered.
A comparable trend is also observed for $^{160,161}$Re, where the inclusion of shape coexistence moderately improves the agreement with the measured half-lives.
For other isotopes, such as $^{176,177}$Tl, however, the inclusion of shape coexistence leads to slightly larger discrepancies from experiment.
The noticeable variations in half-lives are generally found for nuclei in which either the parent or the daughter nucleus exhibits a relatively large energy difference between the prolate and oblate minima (i.e., large $|\Delta E_{\rm tot}|$, see Table~\ref{tab2}), although there are some exceptions such as $^{170}$Au.

\begin{figure}[t]
\begin{center}
\includegraphics[width=0.45\columnwidth]{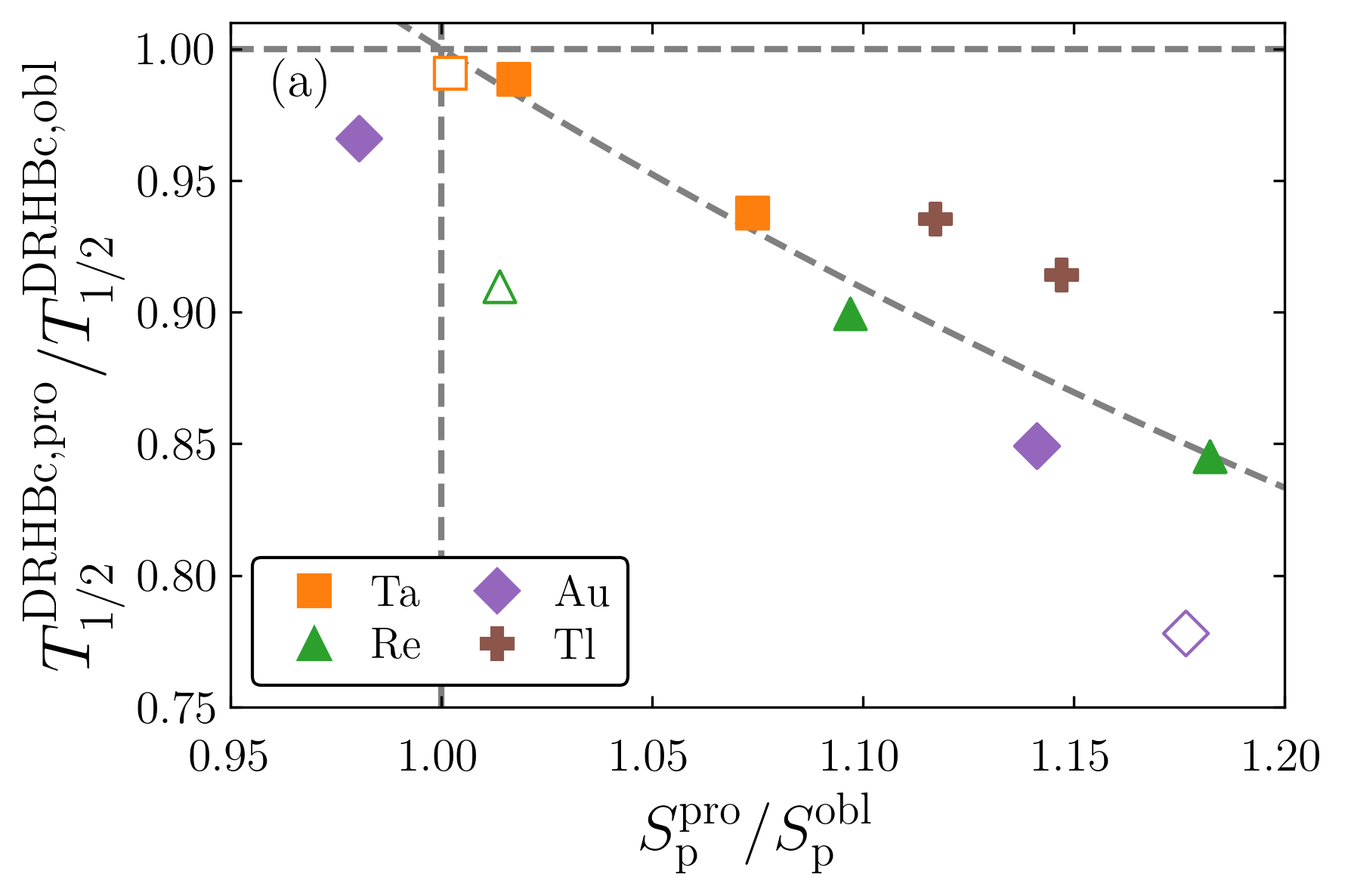}
\includegraphics[width=0.45\columnwidth]{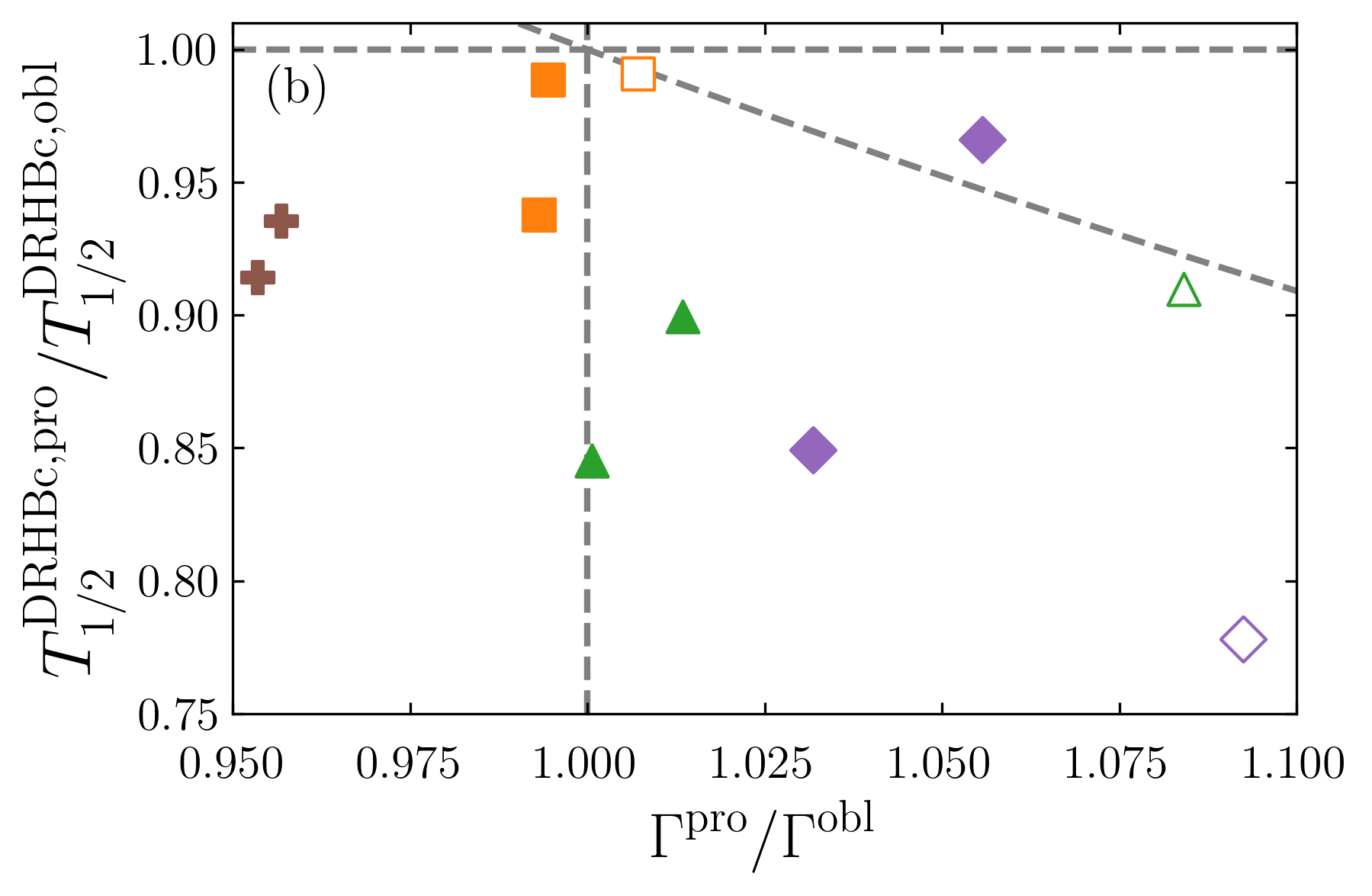}
\includegraphics[width=0.45\columnwidth]{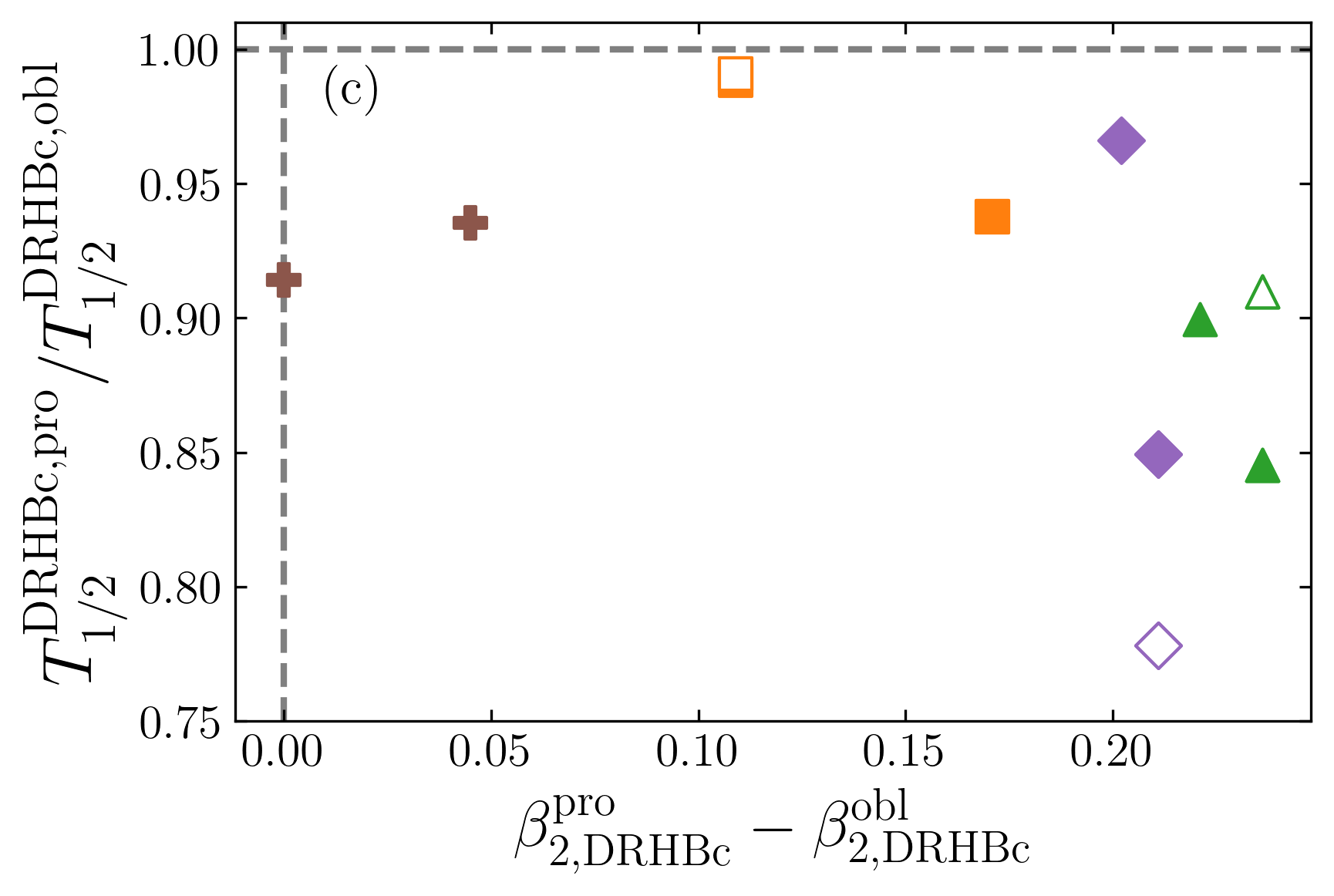}
\includegraphics[width=0.45\columnwidth]{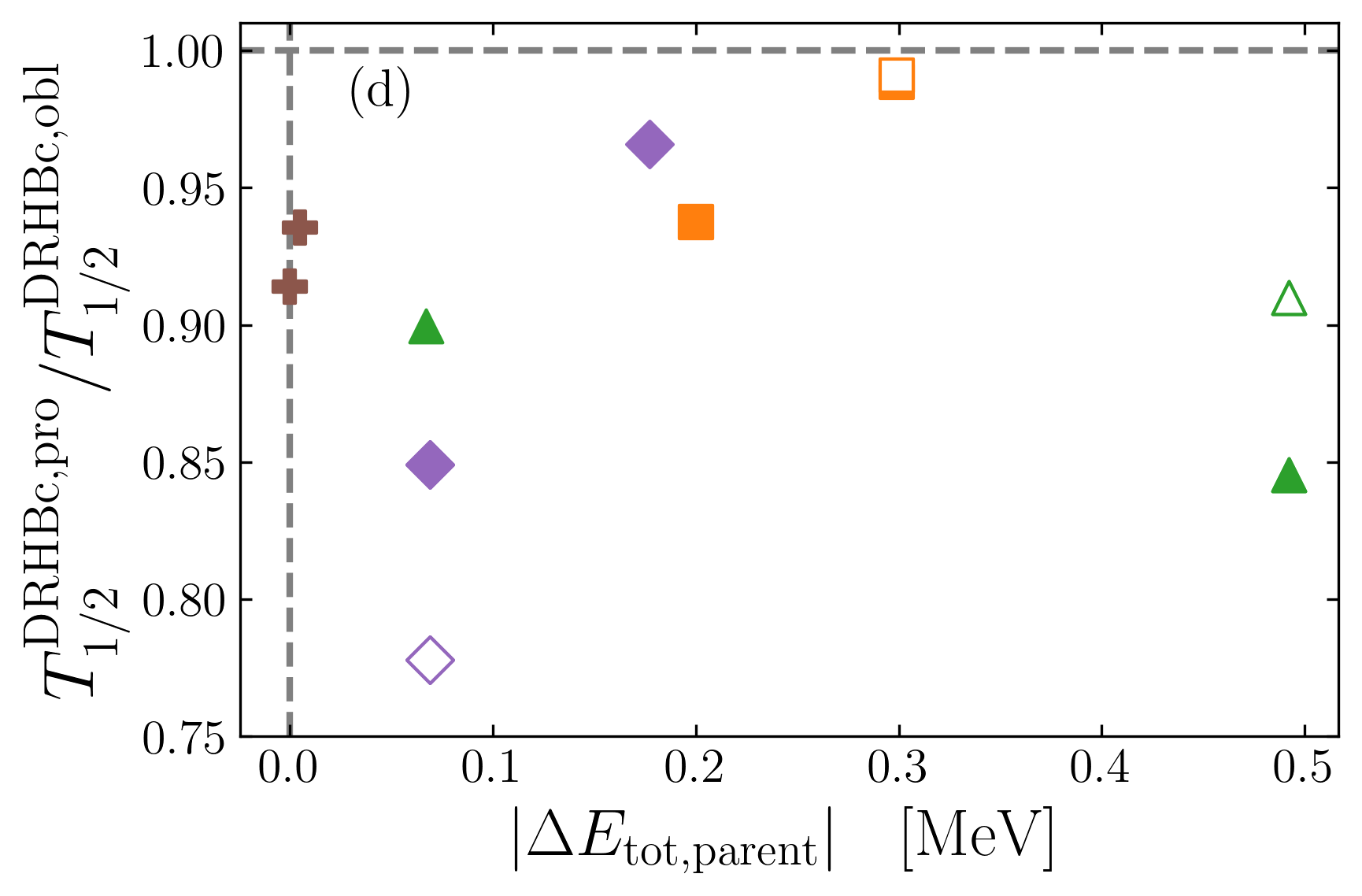}
\includegraphics[width=0.45\columnwidth]{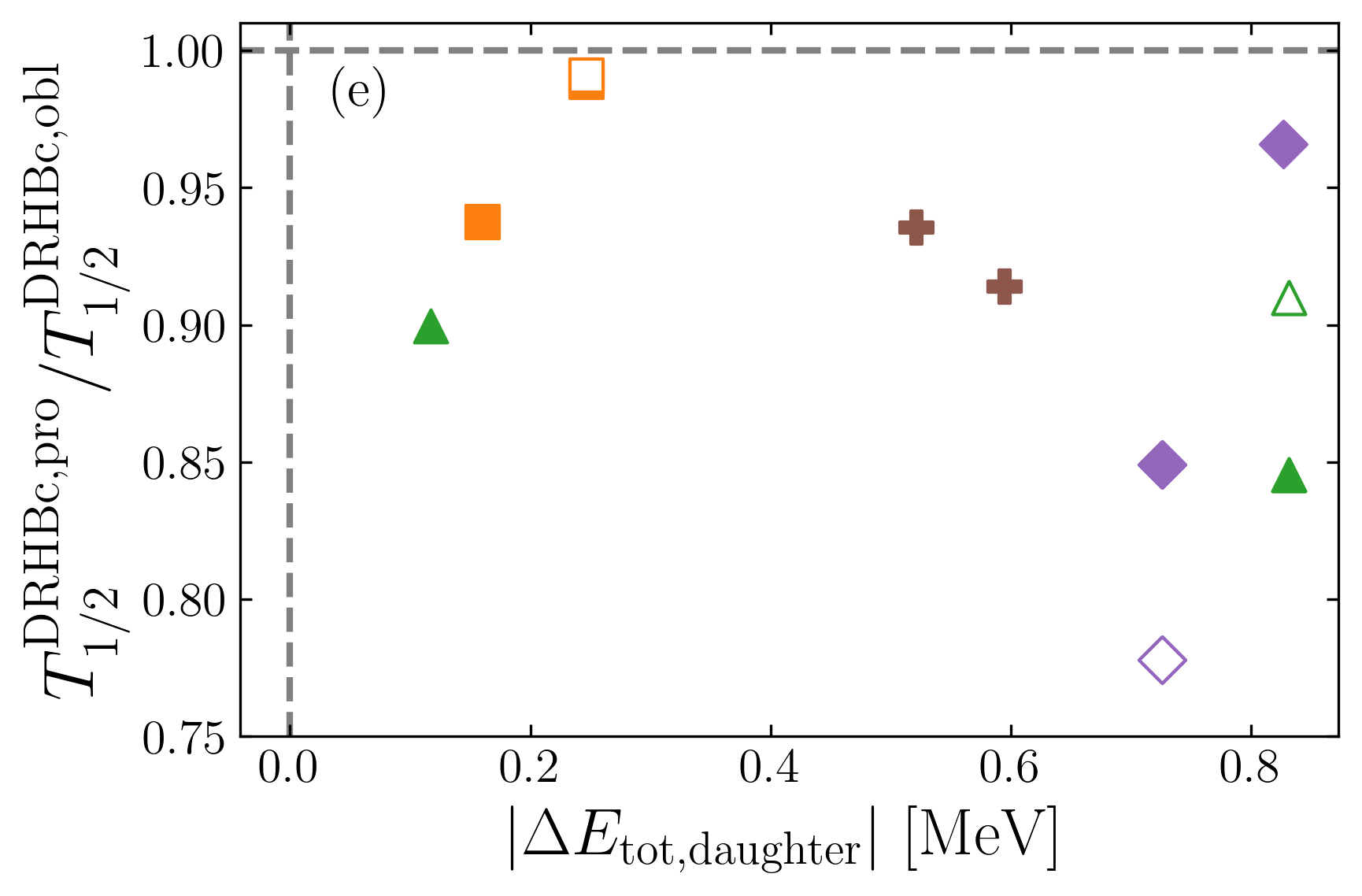}
\caption{
Comparison of the calculated half-life ratio $T^{\rm pro}_{1/2}/T^{\rm obl}_{1/2}$ with nuclear-structure quantities: 
(a) the spectroscopic factor ratio $S_{\rm p}^{\rm pro}/S^{\rm obl}_{\rm p}$, 
(b) the decay width ratio $\Gamma^{\rm pro}/\Gamma^{\rm obl}$, 
(c) the difference in deformation $\beta^{\rm pro}_{2,{\rm DRHBc}} - \beta^{\rm obl}_{2,{\rm DRHBc}}$, 
(d) the total-energy difference of the parent nuclei $|\Delta E_{\rm tot,parent}|$, 
and (e) that of the daughter nuclei $|\Delta E_{\rm tot,daughter}|$. 
For $^{177m}$Tl, the expected inverse relation between the half-life and the spectroscopic factor is indeed valid, since $T^{\rm DRHBc,pro}_{1/2} / T^{\rm DRHBc,obl}_{1/2}=0.607$ and $S_{\rm p}^{\rm pro} / S_{\rm p}^{\rm obl}=1.72$.
} \label{fig6} 
\end{center}
\end{figure}

To further quantify which nuclear-structure quantities govern such variations in half-lives, we next examine their possible correlations.
For nuclei exhibiting shape coexistence, both the prolate and oblate minima are taken into account as independent configurations when evaluating the correlations.
Figure~\ref{fig6} displays possible correlations between the ratio of the half-lives in the prolate and oblate minima and several nuclear-structure quantities.
In particular, panels (a) and (b) compare the influence of the spectroscopic factor and the decay width, respectively. 
It is clearly seen that the variation of the spectroscopic factor exhibits a stronger correlation with the half-life ratio than that of the decay width. 
This trend is further supported by the auxiliary guide line $y=1/x$, to which the data in Fig.~\ref{fig6}(a) are closer. 
In contrast, the deviations in half-lives show little systematic dependence on either the differences of the deformations $\beta^{\rm pro}_{2,{\rm DRHBc}} - \beta^{\rm obl}_{2,{\rm DRHBc}}$ [see Fig.~\ref{fig6}(c)] or the total-energy differences of the parent nuclei $|\Delta E_{\rm tot,parent}|$ [see Fig.~\ref{fig6}(d)].
Likewise, the total-energy differences of the daughter nuclei $|\Delta E_{\rm tot,daughter}|$ [see Fig.~\ref{fig6}(e)] exhibit no clear correlation with the calculated half-lives.
These results indicate that, within the present framework, the spectroscopic factor plays the more dominant role in determining the calculated half-lives than other quantities such as deformation and total-energy differences.

\begin{table*}[h!]
\centering
\caption{Calculated half-lives with DRHBc and RCHB for nuclei exhibiting shape coexistence. Shape coexistence is identified by the presence of additional minima with $|\Delta E_{\rm tot}| \leq 0.6$ MeV. The total-energy differences are listed separately for parent and daughter nuclei. In the case of Tl isotopes, the daughter nuclei exhibit shape coexistence. Columns “P”, “D”, “pro”, and “obl” denote the parent, daughter, prolate, and oblate minimum, respectively.} \label{tab2}
\begin{tabular}{|c|c|c|c|c|c|c|c|c|c|c|c|c|c|c|c|}
\hline
Parent & \multicolumn{4}{c|}{$\beta_{2,{\rm DRHBc}}$} & $Q$ & \multirow{2}{*}{Orbital} & \multicolumn{3}{c|}{$S_{\rm p}$} & \multicolumn{4}{c|}{$\log_{10} T_{1/2}$ [s]} & \multicolumn{2}{c|}{$|\Delta E_{\rm tot}|$ [MeV]} \\ 
\cline{2-5} \cline{8-10} \cline{11-14} \cline{15-16}
nucleus & P$^{\rm pro}$ & D$^{\rm pro}$ & P$^{\rm obl}$ & D$^{\rm obl}$ & [MeV] &  & DRHBc$^{\rm pro}$ & DRHBc$^{\rm obl}$ & RCHB & Exp & DRHBc$^{\rm pro}$ & DRHBc$^{\rm obl}$ & RCHB & \phantom{aaa}P\phantom{aaa} & D \\ 
\hline
$^{\phantom{m}156}$Ta & 0.048 &  0.052 & $-$0.061 & $-$0.059 &   1.030 & $2d_{3/2}$ &  0.533 &  0.524 &  0.534 & $-$0.609 & $-$0.800 & $-$0.795 & $-$0.794 &  0.299 &  0.247 \\
$^{156m}$Ta & 0.048 &  0.052 & $-$0.061 & $-$0.059 &  1.127 & $1h_{11/2}$ &  0.481 &  0.480 &  0.484 & \phantom{$-$}0.930 & \phantom{$-$}0.950 & \phantom{$-$}0.954 &  \phantom{$-$}0.953 &  0.299 &  0.247 \\
$^{\phantom{m}157}$Ta & 0.089 &  0.076 & $-$0.082 & $-$0.083 &   0.947 & $3s_{1/2}$ &  0.689 &  0.650 &  0.713 & $-$0.523 & $-$0.482 & $-$0.454 & $-$0.478 &  0.200 &  0.160 \\
$^{\phantom{m}160}$Re &  0.118 &  0.118 & $-$0.103 & $-$0.106 &  1.285 & $2d_{3/2}$ &  0.475 &  0.433 &  0.448 & $-$3.060 & $-$3.381 & $-$3.335 & $-$3.320 &  0.067 &  0.117 \\
$^{\phantom{m}161}$Re &  0.131 &  0.134 & $-$0.105 & $-$0.124 &  1.214 & $3s_{1/2}$ &  0.655 &  0.554 &  0.677 & $-$3.357 & $-$3.566 & $-$3.493 & $-$3.542 &  0.492 &  0.831 \\
$^{161m}$Re &  0.131 &  0.134 & $-$0.105 & $-$0.124 &  1.338 & $1h_{11/2}$ &  0.368 &  0.363 &  0.370 & $-$0.650 & $-$0.890 & $-$0.849 & $-$0.842 &  0.492 &  0.831 \\
$^{\phantom{m}170}$Au &  0.104 &  0.136 & $-$0.098 & $-$0.102 &  1.488 & $2d_{3/2}$ &  0.252 &  0.257 &  0.270 & $-$3.493 & $-$4.075 & $-$4.060 & $-$4.048 &  0.177 &  0.827 \\
$^{170m}$Au &  0.104 &  0.136 & $-$0.098 & $-$0.102 &  1.770 & $1h_{11/2}$ &  0.185 &  0.149 &  0.126 & $-$2.980 & $-$3.374 & $-$3.237 & $-$3.142 &  0.177 &  0.827 \\
$^{\phantom{m}171}$Au &  0.108 &  0.135 & $-$0.103 & $-$0.110 &  1.464 & $3s_{1/2}$ &  0.525 &  0.460 &  0.647 & $-$4.611 & $-$4.996 & $-$4.925 & $-$5.033 &  0.069 &  0.726 \\
$^{171m}$Au &  0.108 &  0.135 & $-$0.103 & $-$0.110 &  1.719 & $1h_{11/2}$ &  0.180 &  0.153 &  0.124 & $-$2.654 & $-$2.997 & $-$2.888 & $-$2.770 &  0.069 &  0.726 \\
$^{\phantom{m}176}$Tl & 0.013 &  0.056 & $-$0.032 & $-$0.097 & 1.282 & $3s_{1/2}$ &  0.524 &  0.469 &  0.581 & $-$2.284 & $-$2.498 & $-$2.469 & $-$2.531 &  0.005 &  0.521 \\
$^{\phantom{m}177}$Tl & 0.000 &  0.056 & \phantom{$-$}0.000 & $-$0.101 & 1.180 & $3s_{1/2}$ &  0.530 &  0.462 &  0.589 & $-$1.174 & $-$1.243 & $-$1.204 & $-$1.277 &  0.000 &  0.595 \\
$^{177m}$Tl & 0.000 & 0.056 & \phantom{$-$}0.000 & $-$0.101 &  1.984 & $1h_{11/2}$ &  0.050 &  0.029 &  0.041 & $-$3.402 & $-$3.785 & $-$3.568 & $-$3.684 &  0.000 &  0.595 \\
\hline
\end{tabular}
\end{table*}

\section{Summary and Discussion} \label{summary_section}
In this work, we investigate the half-lives of one-proton emitters of $71 \leq Z \leq 83$ odd-$Z$ nuclei by employing the WKB approximation with nuclear potentials derived from the DRHBc. For comparison, we also perform calculations with RCHB. The calculated half-lives are then systematically compared with the available experimental data.

Our main findings can be summarized as follows.  
First, both DRHBc and RCHB calculations systematically underestimate the measured half-lives, corresponding to an overestimation of the proton-emission probability. 
This discrepancy suggests that the decay width $\Gamma$ may be biased toward larger values within the present framework.
Second, the spectroscopic factor $S_{\rm p}$ is found to exert a strong influence on the variation of the half-lives. In contrast, the contribution from the decay width is relatively modest within the limited deformation range ($|\beta_{2,{\rm DRHBc}}| \lesssim 0.15$). However, since the decay width depends exponentially on the $Q$ value, the uncertainty of the $Q$ value remains the most critical factor for the overall accuracy of the half-life calculations.  
Third, in nuclei exhibiting shape coexistence, the inclusion of multiple configurations can either improve or worsen the agreement with experiment depending on the specific isotope. For example, in $^{177m}$Tl the deviation from experiment becomes larger, while in $^{170,170m,171,171m}$Au and $^{160,161}$Re the agreement is moderately improved. This indicates that the role of shape coexistence is highly nucleus-dependent, reflecting the delicate balance between the deformation dependence of the spectroscopic factor and the decay width (tunneling probability).

Despite these findings, several limitations of the present framework should be noted.
First, the DRHBc and RCHB calculations are based on the quasiparticle mean-field approximation, in which particle number is conserved only on average and no explicit particle-number projection is performed. 
The self-consistent mean-field potentials used in the decay calculation are derived from quasiparticle states rather than exact particle-number eigenstates, which may lead to deviations in occupation probabilities near the Fermi surface and omit beyond-mean-field correlations (e.g., particle-number restoration or configuration mixing).
Second, although continuum effects are incorporated in both DRHBc and RCHB (through the Dirac Woods–Saxon basis and by solving the relativistic Hartree-Bogoliubov equations in coordinate space), no explicit resonance solutions satisfying outgoing-wave boundary conditions are included.
Third, the decay width is evaluated within the semiclassical WKB approximation, which may introduce systematic uncertainties in the tunneling probability.
In addition, the present treatment does not include angular-momentum projection or a full decomposition into partial decay widths, which may further affect the quantitative evaluation of the decay width.
Since the calculated half-lives are systematically smaller than the experimental values and no clear systematic bias is identified in the spectroscopic factor, an overestimation of the decay width may account for the observed discrepancy.
The systematic uncertainties inherent in the WKB approximation for tunneling probabilities are discussed in Ref.~\cite{DONG20111}.
Furthermore, in nuclei exhibiting shape coexistence, the present treatment relies on a simple averaging of near-degenerate minima without explicit configuration mixing, which may introduce additional uncertainties in the predicted half-lives despite the small energy difference between the minima.

In conclusion, the present results demonstrate that while the spectroscopic factor is the dominant structural quantity in determining proton emission half-lives within DRHBc, the present WKB framework still shows systematic deviations from experiment. 
This suggests that more advanced treatments of the tunneling process, beyond the semiclassical WKB approximation, will be required to reduce these discrepancies \cite{Fan2025}.
Since the tunneling probability depends exponentially on the potential barrier, such refinements are expected to significantly improve agreement with experiment. 
Nevertheless, as exemplified by the case of $^{185}$Bi ($Z=83$), which lies just above the proton shell closure at $Z = 82$, the mean-field description itself may have intrinsic limitations in properly describing the proton–daughter nucleus system.
Therefore, continuous improvement of self-consistent mean-field theories, for example by including beyond-mean-field effects, together with a more sophisticated treatment of the decay dynamics, will be essential for achieving reliable predictions of exotic decay modes and for deepening our understanding of nuclear properties near the proton drip line.


\section*{ACKNOWLEDGMENTS}
Helpful discussions with the members of the DRHBc Mass Table Collaboration are greatly appreciated.
This project was started while Y.K. was visiting Pusan National University in 2023.
Y.C. was supported in part by the National Key R\&D Program of China (2022YFA1602401) and the National Natural Science Foundation of China (No. 12335009 \& 12435010).
C.-H.L. was supported by the National Research Foundation of Korea (NRF) grant funded by the Korea government (MSIT) (No. 2023R1A2C1005398).
Y.K. was supported in part by the Institute for Basic Science (IBS-R031-D1).
This work was supported by the National Supercomputing Center with supercomputing resources including technical support (KSC-2023-CRE-0521).

\bibliographystyle{aip}

\pagebreak

\end{document}